\UseRawInputEncoding
\documentclass[%
 reprint,
 amsmath,amssymb,
 aps,
]{revtex4-2}

\usepackage{graphicx}
\usepackage{dcolumn}
\usepackage{bm}

\usepackage[symbol]{footmisc}
\usepackage{graphicx}
\usepackage{soul,color}
\usepackage{physics}
\usepackage{lineno}
\usepackage[english]{babel}
\usepackage{fancyhdr}
\usepackage{lastpage}
\usepackage{float}
\usepackage[normalem]{ulem}
\usepackage{tikz}
\usepackage{float}
\usetikzlibrary{quantikz}
 
\begin{document}


\renewcommand{\thefootnote}{\fnsymbol{footnote}}
\newcommand{\ttH}{\ensuremath{t\bar{t}H}\,}
\newcommand{\Hmumu}{\ensuremath{H\rightarrow\mu^{+}\mu^{-}}\,}
\newcommand{\Hgamgam}{\ensuremath{H\rightarrow\gamma\gamma}\,}

\vspace{0.5cm}

\title{Application of Quantum Machine Learning using the Quantum Kernel Algorithm on High Energy Physics Analysis at the LHC}

\author{Sau Lan Wu}
\email{sau.lan.wu@cern.ch}
\affiliation{Department of Physics, University of Wisconsin, Madison WI, USA}

\author{Shaojun Sun}
\affiliation{Department of Physics, University of Wisconsin, Madison WI, USA}

\author{Wen Guan}
\affiliation{Department of Physics, University of Wisconsin, Madison WI, USA}

\author{Chen Zhou}
\affiliation{Department of Physics, University of Wisconsin, Madison WI, USA}

\author{Jay Chan}
\affiliation{Department of Physics, University of Wisconsin, Madison WI, USA}

\author{Chi Lung Cheng}
\affiliation{Department of Physics, University of Wisconsin, Madison WI, USA}

\author{Tuan Pham}
\affiliation{Department of Physics, University of Wisconsin, Madison WI, USA}

\author{Yan Qian}
\affiliation{Department of Physics, University of Wisconsin, Madison WI, USA}

\author{Alex Zeng Wang}
\affiliation{Department of Physics, University of Wisconsin, Madison WI, USA}

\author{Rui Zhang}
\affiliation{Department of Physics, University of Wisconsin, Madison WI, USA}

\author{Miron Livny}
\affiliation{Department of Computer Sciences, University of Wisconsin, Madison WI, USA}

\author{Jennifer Glick}
\affiliation{IBM Quantum, T.J. Watson Research Center, Yorktown Heights, NY, USA }

\author{Panagiotis Kl. Barkoutsos}
\affiliation{IBM Quantum, Zurich Research Laboratory, R\"̈uschlikon, Switzerland}

\author{Stefan Woerner}
\affiliation{IBM Quantum, Zurich Research Laboratory, R\"̈uschlikon, Switzerland}

\author{Ivano Tavernelli}
\affiliation{IBM Quantum, Zurich Research Laboratory, R\"̈uschlikon, Switzerland}

\author{Federico Carminati}
\affiliation{CERN Quantum Technology Initiative, IT Department, CERN, Geneva, Switzerland}

\author{Alberto Di Meglio}
\affiliation{CERN Quantum Technology Initiative, IT Department, CERN, Geneva, Switzerland}

\author{Andy C. Y. Li}
\affiliation{Quantum Institute, Fermi National Accelerator Laboratory, Batavia, IL, USA }

\author{Joseph Lykken}
\affiliation{Quantum Institute, Fermi National Accelerator Laboratory, Batavia, IL, USA }

\author{Panagiotis Spentzouris}
\affiliation{Quantum Institute, Fermi National Accelerator Laboratory, Batavia, IL, USA }

\author{Samuel Yen-Chi Chen}
\affiliation{Computational Science Initiative, Brookhaven National Laboratory, Upton, NY, USA }

\author{Shinjae Yoo}
\affiliation{Computational Science Initiative, Brookhaven National Laboratory, Upton, NY, USA }

\author{Tzu-Chieh Wei}
\affiliation{C.N. Yang Institute for Theoretical Physics, State University of New York at Stony Brook, Stony Brook, NY, USA }

\begin{abstract}
Quantum machine learning could possibly become a valuable alternative to classical machine learning for applications in High Energy Physics by offering computational speed-ups.
In this study, we employ a support vector machine with a quantum kernel estimator (QSVM-Kernel method) to a recent LHC
flagship physics analysis: \ttH\ (Higgs boson production in association with a top quark pair). 
In our quantum simulation study using up to 20 qubits and up to 50000 events, the QSVM-Kernel method performs as well as its classical counterparts in three different platforms from Google Tensorflow Quantum, IBM Quantum and Amazon Braket. 
Additionally, using 15 qubits and 100 events, the application of the QSVM-Kernel method on the IBM superconducting quantum hardware approaches the performance of a noiseless quantum simulator.
Our study confirms that the QSVM-Kernel method can use the large
dimensionality of the quantum Hilbert space to replace the classical feature space in realistic physics datasets. 
\end{abstract}

\maketitle

\section{INTRODUCTION}

In 2012, the ATLAS and CMS experiments discovered the Higgs boson~\cite{atlashiggs,cmshiggs} using proton-proton collision data at the Large Hadron Collider (LHC). This discovery completed the fundamental particle spectrum of the Standard Model and was a major achievement in High Energy Physics (HEP).
As the LHC experiments enter the post-Higgs discovery era, physicists strive to refine the understanding of the Standard Model and pursue new physics beyond the Standard Model.
Machine learning has become one of the most powerful tools for exploring the full physics potential of the huge amount of data collected by the LHC experiments. 
In HEP, the important usage of machine learning techniques~\cite{ml4hep1,ml4hep2,ml4hep3,ml4hep4,ml4hep5} includes simulation, event reconstruction and data analyses. 
As an example, the ATLAS and CMS experiments utilized supervised machine learning algorithms in the recent data analyses to achieve the observation of the \ttH\ processs~\cite{atlastth,cmstth}, a very rare Higgs production mode.

Quantum machine learning, where machine learning is performed using quantum algorithms, has the potential to improve the computational complexity of classical machine learning algorithms and obtain computational speed-ups when being executed on quantum computers~\cite{qml}. 
Some of the quantum algorithms will benefit from exponential improvements in speed.
For classification problems, it may also lead to better separation power than classical machine learning. 
A recent result~\cite{qml-pa} demonstrates a significant advantage in prediction accuracy for a quantum algorithm over some classical algorithms on engineered datasets.
A key component of quantum machine learning algorithms is exploiting the high dimensional quantum state space through the actions of superposition and entanglement.
With the progress of quantum technologies, quantum machine learning could possibly become a valuable alternative to classical machine learning for processing big data (including simulation, reconstruction and analyses) in High Energy Physics~\cite{qml-hep}. Recent projections from IBM~\cite{ibm-roadmap}, Google~\cite{google-roadmap}, and IonQ~\cite{ionq-roadmap} suggest that  quantum computers with thousands of qubits capable of performing practical computational tasks may become available within the next ten years. This coincides with the High Luminosity upgrades of the LHC (HL-LHC)~\cite{Apollinari-2017lan}, bringing a massive increase in not only the amount of collected physics data, but also the computational power needed to process that data. Present research on quantum machine learning algorithms could be implemented on those future devices, thereby ensuring the timely exploitation of quantum advantages for physics applications, and possibly even contributing to the discovery of new physics.

The challenges we face for the future include working with large numbers of events in the millions, applying large numbers of qubits in the thousands to obtain stellar performance, and achieving excellent computational speeds. It is the purpose of this publication to inch one step closer to conquering those challenges, even under the present limitations of existing quantum computer technology.

Previous studies~\cite{qml4hep0, qml4hep1,qml4hep2} have investigated quantum annealing or quantum classifiers trained with variational circuits. A quantum machine learning algorithm, the support vector machine with a quantum kernel estimator (QSVM-Kernel), was proposed to solve classification problems~\cite{qsvmv,qsvmm}. 
This algorithm was experimentally implemented with 2 qubits on a superconducting quantum computer and found to be accurate for artificial datasets~\cite{qsvmv}.
QSVM-Kernel leverages the quantum state space as a direct representation of the feature space, which can give rise to kernel functions that are hard to evaluate classically~\cite{qsvmv,forrelation}.
However, any potential quantum advantage does not lie solely in the high dimensionality of the quantum state space, as it is well known that classical kernels can map into feature space of arbitrary high dimensions.
Rather, a path towards quantum advantage is found in the computational complexity of the quantum circuits used to compute the quantum kernel. 
Namely, these circuits must be hard to estimate classically. A recent result~\cite{qsvm-dlp} establishes an exponential quantum speedup for QSVM-Kernel using a fault-tolerant quantum computer to estimate the kernel function for a classically hard classification problem. Although this particular problem was not practically motivated, the result rigorously formalizes the intuition that quantum feature maps can identify patterns classical machines are unable to capture. From this foundation, quantum feature maps can be designed and tested on practical data sets, potentially leading to better classification results than classical feature maps and kernels. 

It is therefore interesting to study the QSVM-Kernel algorithm with a larger number of qubits and evaluate its performance on real-world datasets. 
In our study, we successfully employ the QSVM-Kernel algorithm in the \ttH\ analysis, 
a recent LHC flagship physics analysis, using up to 20 qubits on quantum computer simulators and up to 15 qubits on quantum computer hardware. 
Furthermore, we compare the classification result of the QSVM-Kernel algorithm to a few popular classical algorithms that are commonly used by the LHC experiments.

\section{\ttH\ Physics Analysis at the LHC}
The observation of \ttH\ (Higgs boson production in association with a top quark pair) by the ATLAS and CMS experiments~\cite{atlastth,cmstth} was one of the LHC flagship physics results following the Higgs boson discovery. 
It established a direct observation of the Higgs boson's interaction with the top quark, the heaviest known fundamental particle.
Study of the Higgs-top interaction may provide crucial test for the Higgs mechanism of the Standard Model and essential clues for new physics beyond the Standard Model.
Due to the small \ttH\ production rate at the LHC, its observation was highly challenging. 
To achieve the desired sensitivities to $\ttH$ production, the ATLAS and CMS collaborations combined results from a number of decay channels. 
The physics analyses in many of these channels utilize machine learning techniques.
For example, classifiers based on machine learning are constructed to analyze kinematic variables of the collision events and distinguish between signal and background. 

\begin{figure}[htb]
\begin{center}
\includegraphics[width=1.7in]{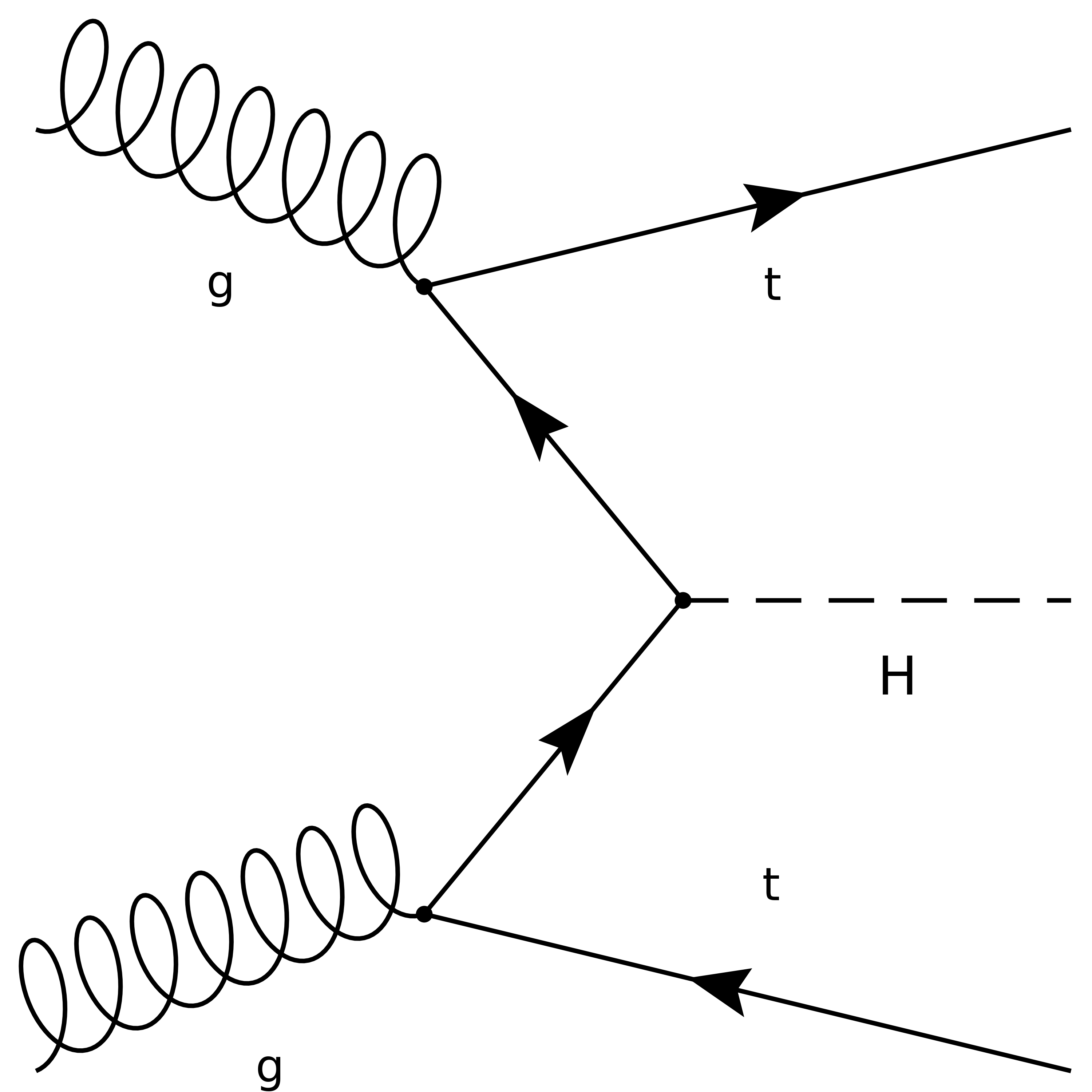}\\
(a)\\
\includegraphics[width=1.7in]{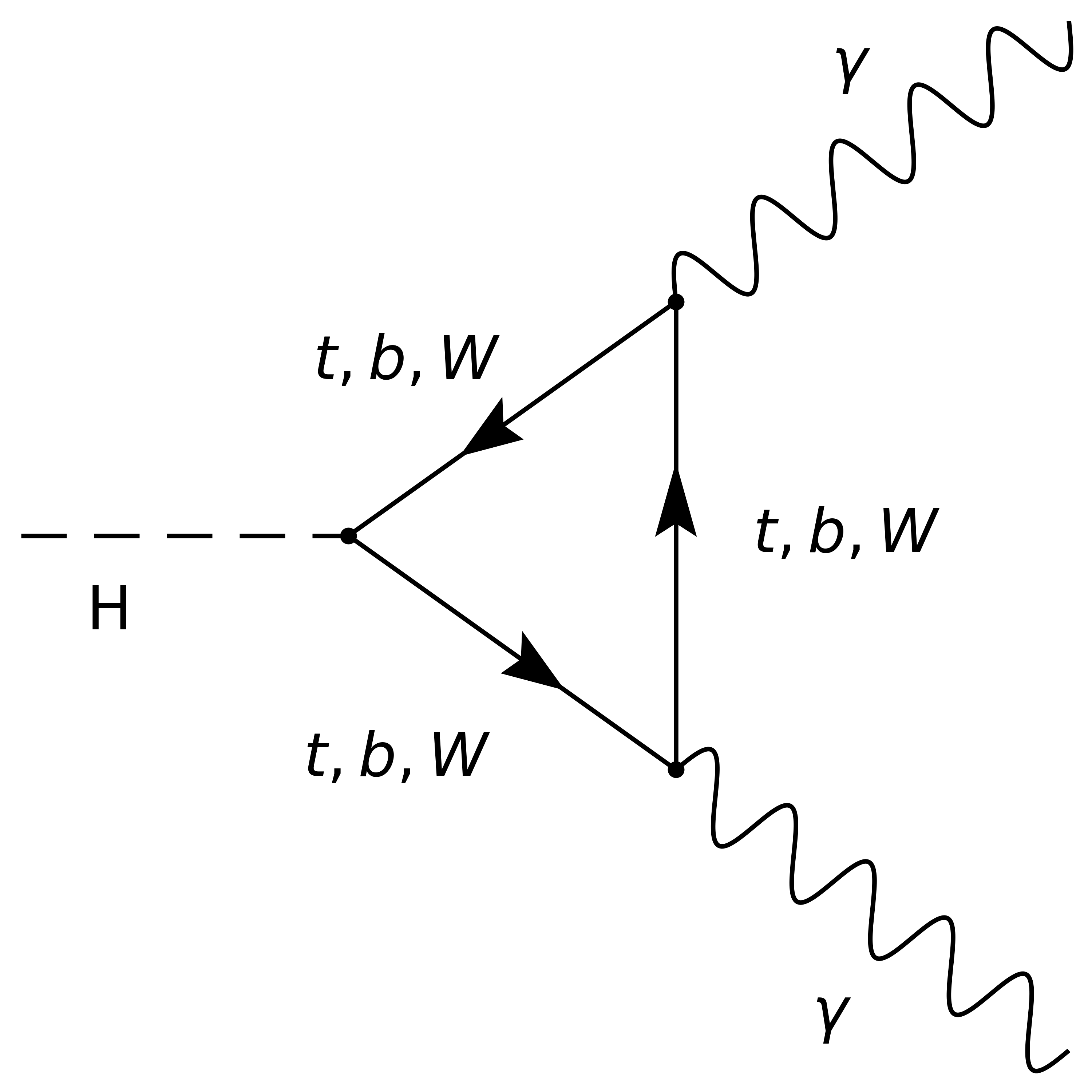}\\
(b)\\
\includegraphics[width=1.7in]{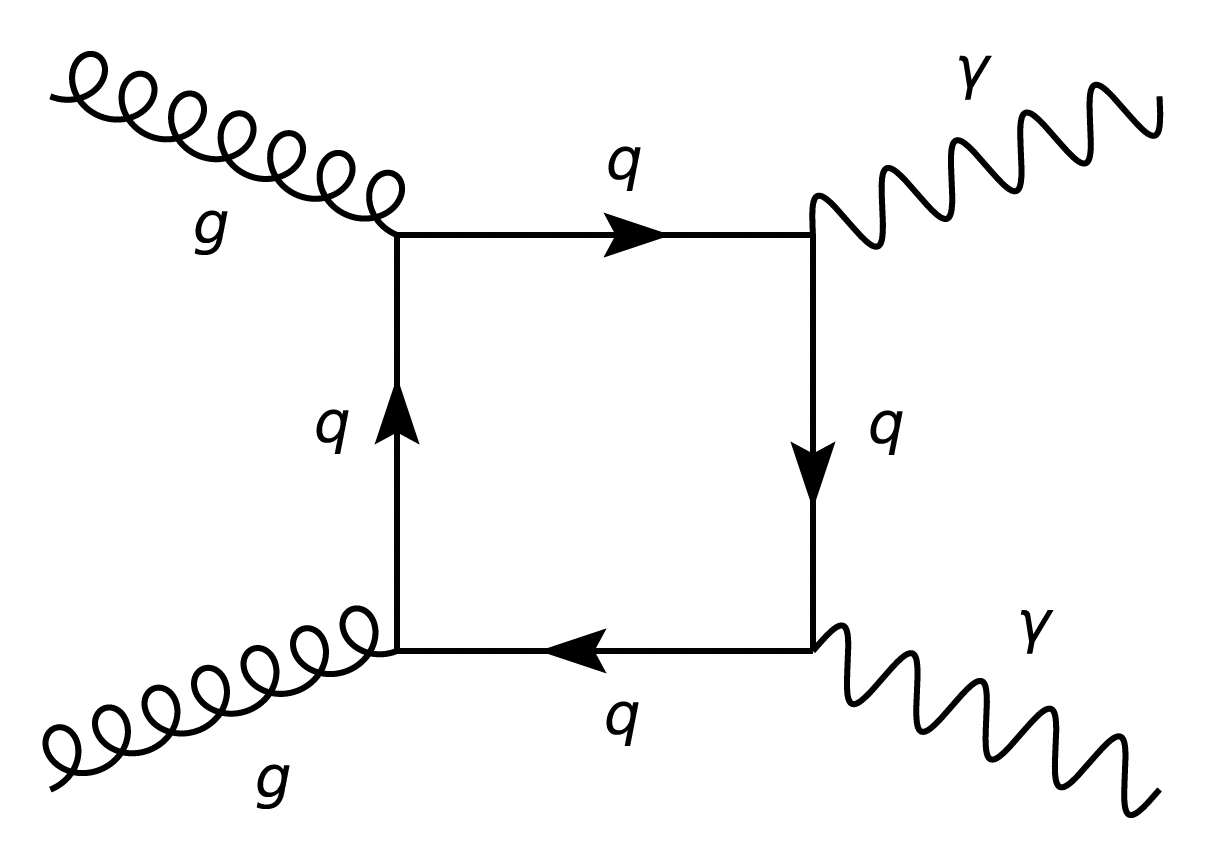}\\
(c)\\
\end{center}
\caption{Representative Feynman diagrams for (a) \ttH\ production, (b) \Hgamgam\ decay, and (c) non-resonant two-photon production. 
In these diagrams, $H$ denotes a Higgs boson, $g$ denotes a gluon, $q$ denotes a quark, $t$ denotes a top quark, $b$ denotes a bottom quark, $W$ denotes a W boson, and $\gamma$ denotes a photon.}
\label{fig_feynman_diagrams}
\end{figure}

In our study, we focus on an important \ttH\ analysis channel where the Higgs boson decays into two photons (\Hgamgam) and the two top quarks decay into hadrons.
The dominant background in this analysis channel is non-resonant two-photon production.
See Figure~\ref{fig_feynman_diagrams} for representative Feynman diagrams for \ttH\ production, \Hgamgam\ decay, and non-resonant two-photon production.
\\

\section{Method}
\label{Method}
The support vector machine (SVM)~\cite{svm,svm2} is one of the most commonly used supervised machine learning algorithms for data classification. 
Here for a data event, $\vec{x}$ denotes the vector of its input features and $y \in \{0,1\}$ denotes its class label (0 for background and 1 for signal).
The SVM algorithm maps $\vec{x}$ into a higher dimensional feature space, 
where it measures the similarity between any two data events (denoted as ``kernel entry'', $k(\vec{x}_i, \vec{x}_j)$).
The SVM algorithm then optimizes a hyperplane that separates signal events from background events and classifies a new data event $\vec{x}'$ by
\begin{equation}
        y' = sgn (\sum_{i=1}^{t} \alpha_i y_i k(\vec{x}_i, \vec{x}') + b)      
\end{equation}
where $sgn$ is the sign function, $t$ is the size of the training dataset with known labels $y_i$, 
and $(\alpha_i, b)$ defines the separating hyperplane.
Furthermore, a continuous SVM discriminant can be obtained by computing the probabilities of being in the signal class.  
A main limitation of the classical SVM algorithm is that evaluating kernel entries in a large feature space can be computationally expensive.
Three different popular classical kernels are considered to benchmark the performance of the classical SVM method in our study: the linear kernel $\vec{x}_i^T \vec{x}_j$, the polynomial kernel $(\gamma \vec{x}_i^T \vec{x}_j)^d$, and the RBF kernel $\exp(- \gamma \| \vec{x}_i - \vec{x}_j \|^2 )$ 
($\gamma$ and $d$ are hyperparameters). \\

\textbf{A. Quantum Kernel Estimation}\\
A quantum version of the SVM with a quantum kernel estimator (QSVM-Kernel) was introduced in Ref~\cite{qsvmv,qsvmm}, which leverages the quantum state space as a feature space to efficiently compute kernel entries. 
This algorithm maps the classical data event $\vec{x}$ non-linearly to a quantum state of N qubits by applying a quantum feature map circuit $\mathcal{U}_{\Phi(\vec{x})}$ to the initial state $|0^{\otimes N}\rangle$:
\begin{equation}
\ket{\Phi(\vec{x})} = \mathcal{U}_{\Phi(\vec{x})}|0^{\otimes N}\rangle 
\end{equation}
It then calculates the kernel entry for data events $\vec{x_i}$ and $\vec{x_j}$ based on the inner product of their quantum states:
\begin{equation}
    k(\vec{x_i}, \vec{x_j}) = |\langle\Phi(\vec{x_i})|\Phi(\vec{x_j})\rangle|^2 
    = |\langle0^{\otimes N}|\mathcal{U}^{\dag}_{\Phi(\vec{x_i})}\mathcal{U}_{\Phi(\vec{x_j})}|0^{\otimes N}\rangle|^2
\end{equation}
The kernel entry can be evaluated on a quantum computer by measuring the $\mathcal{U}^{\dag}_{\Phi(\vec{x_i})}\mathcal{U}_{\Phi(\vec{x_j})}|0^{\otimes N}\rangle$ state in the computational basis with repeated measurement shots and recording the probability of collapsing the output into the $|0^{\otimes N}\rangle$ state.
In our study, the general design of the quantum circuit for evaluating the kernel entries is inherited from Ref~\cite{qsvmv} and shown in Figure~\ref{fig_qfmap_qvar} (a).\\

\textbf{B. Quantum Feature Map}\\
As suggested in Ref~\cite{qsvmv}, the mathematical properties of the quantum feature map should be complex enough to be hard to simulate on a classical computer, and simple enough to be executable on noisy intermediate-scale quantum computers.  Furthermore, the classification performance of the QSVM-kernel method is highly dependent on the kernel matrix, which is calculated from the quantum states of the feature map. Therefore, choosing a suitable feature map is essential for optimizing the classification performance of the QSVM-kernel method.

The quantum feature map $\mathcal{U}_{\Phi(\vec{x})}$ gives rise to a $2^N$ dimensional feature space (\textit{N} is the number of qubits) that is conjectured to be hard to estimate classically~\cite{forrelation}. 
Following Ref~\cite{qsvmv}, the quantum feature map in our study has two repeated layers:
\begin{equation} 
\mathcal{U}_{\Phi(\vec{x})} = U_{\Phi(\vec{x})} H^{\otimes N} U_{\Phi(\vec{x})} H^{\otimes N} 
\end{equation}
where $H$ is a Hadamard gate and $U_{\Phi(\vec{x_i})}$ is a unitary operator that encodes data from a classical event in its parameters. We analyzed various quantum circuit candidates for the unitary operator $U_{\Phi(\vec{x})}$ and found that the quantum circuit shown in Figure~$\ref{fig_qfmap_qvar}$ (b) performs the best for our $t\bar{t}H$ physics analysis. The chosen quantum circuit is constituted by single-qubit rotation gates ($A$, $B$ and $A'$), as well as two-qubit CNOT entangling gates.
On the $k^{th}$ qubit, given an input feature vector $\vec{x}$,  an $A$ gate rotates the qubit around the $z$ axis of the Bloch sphere by $x_{k}$ (the $k$th element of $\vec{x}$), a $B$ gate rotates the qubit around the $y$ axis by $x_{k}^d$, and an $A'$ gate rotates the qubit around the $z$ axis by $(\frac{x_{k-1} + x_{k}}{2})^d$, 
where $d=3$.
The entangling operations are arranged in an alternating pattern to yield short-depth circuits for execution on noisy intermediate scale quantum computers.
The differences in the quantum feature map between here and Ref~\cite{qsvmv} are mainly the use of $B$ gates, the different parameterization of rotation angles, and the extension to more qubits.
\\

\textbf{C. Separating Hyperplane}\\
In the training phase, the kernel entries are evaluated for all data event pairs of the training sample and then used to find a separating hyperplane.
In the testing phase, the kernel entries are evaluated between a new data event $\vec{x}'$ and each of the data events from the training sample, which are then used to classify the new data event $\vec{x}'$ according to the separating hyperplane. 
For both phases, quantum computers are only used to evaluate the kernel entries. 
Using these kernel entries, the optimization of the separating hyperplane and classification of the new data event are done in classical computers, as for a classical SVM.\\

\begin{figure}[!htb]
    \centering
    \includegraphics[width=\linewidth]{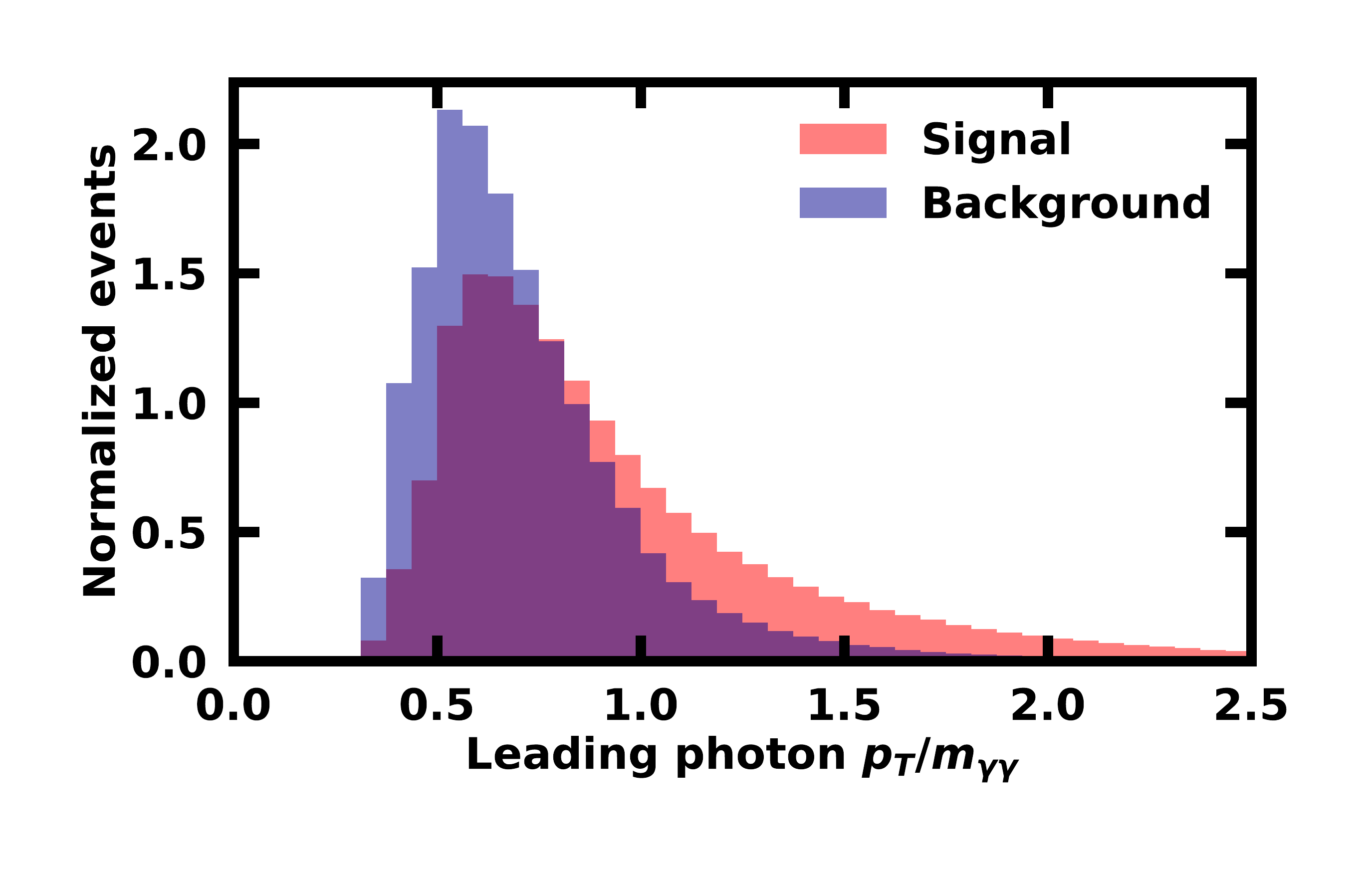}\\
    \textbf{(a)}\\
    \vspace{6pt}      
    \includegraphics[width=\linewidth]{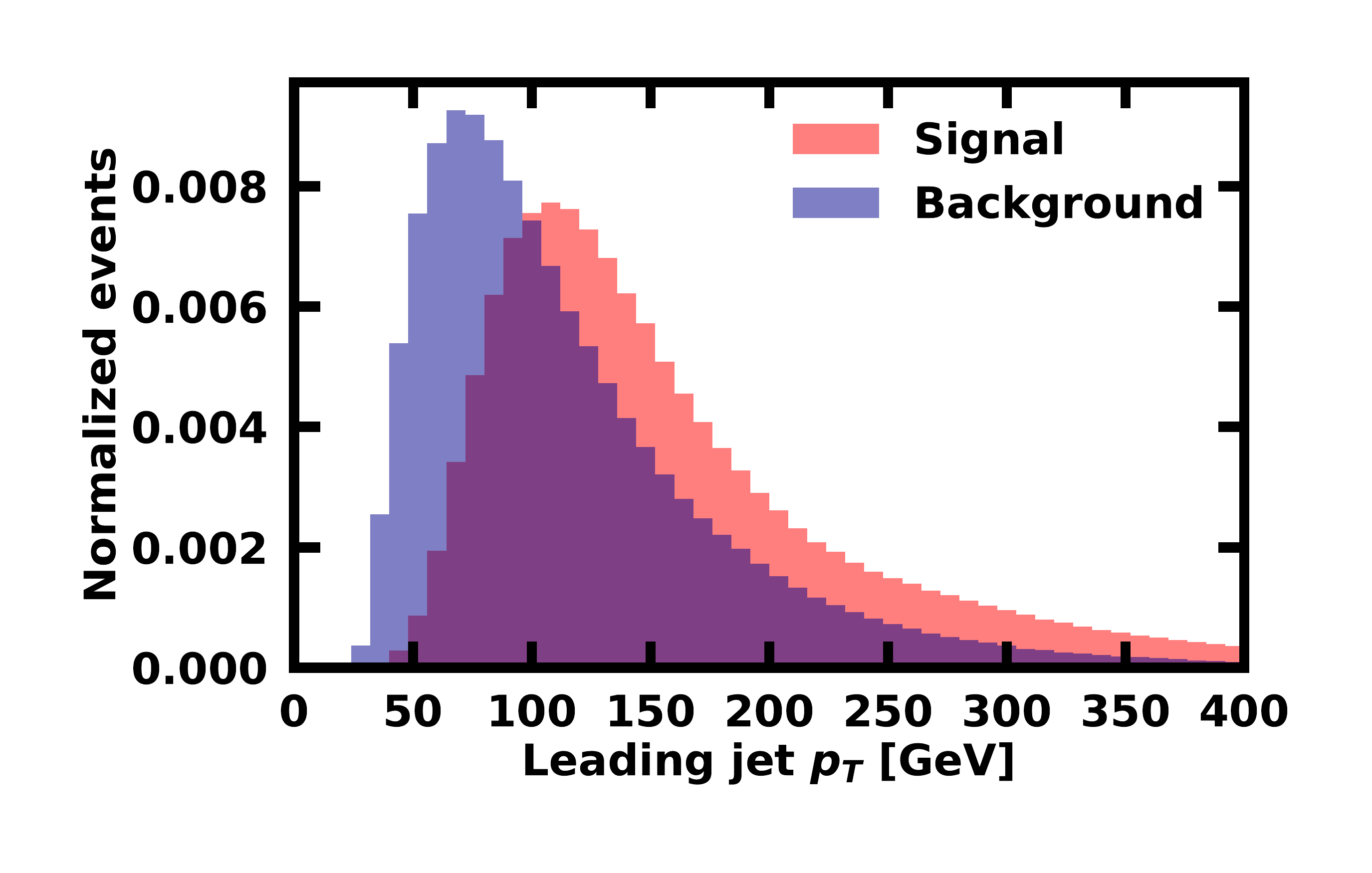}\\
    \textbf{(b)}\\
    \caption{Signal and background distributions for some of the most powerful input variables to the \ttH\ analysis: $\textbf{(a)}$, the transverse momentum of the leading photon divided by the photon pair invariant mass, and $\textbf{(b)}$, the transverse momentum of the leading jet.}
   \label{fig_training_var}
\end{figure}

\begin{figure}[!htb]
    \centering
    \includegraphics[scale=1.0]{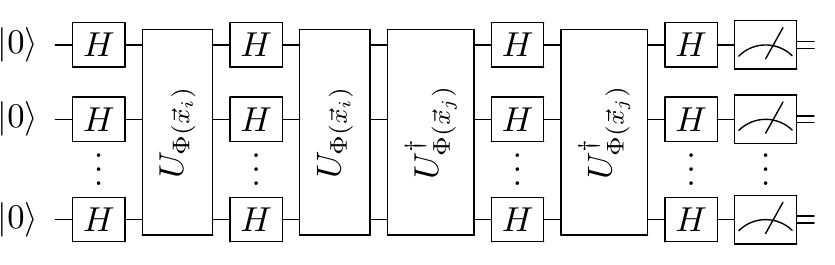}\\
    \textbf{(a)}\\
    \vspace{6pt}      
    \includegraphics[scale=1.0]{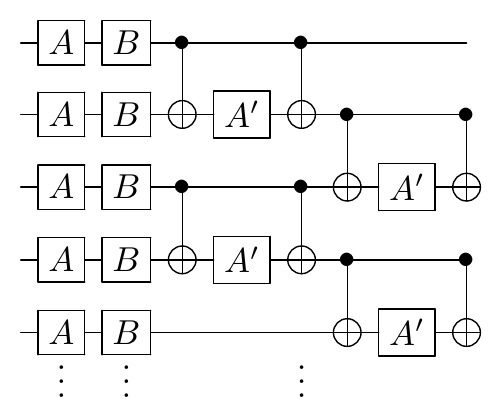}\\
    \textbf{(b)}\\
    \caption{
    \textbf{(a)} Quantum circuit for evaluating the kernel entry for data events $\vec{x_i}$ and $\vec{x_j}$ used in our study. 
    $H$ is a Hadamard gate and $U_{\Phi(\vec{x_i})}$ is a unitary operator that encodes data from a classical event in its parameters.
    \textbf{(b)} Quantum circuit of the unitary operator $U_{\Phi(\vec{x_i})}$. 
    It is constituted by single-qubit rotation gates ($A$, $B$ and $A'$), as well as two-qubit CNOT entangling gates.
    }

   \label{fig_qfmap_qvar}
\end{figure}

\textbf{D. Analysis dataset} \\
In our study for the \ttH\ analysis, the signal and dominant background processes are generated using Madgraph5$\_$aMC@NLO~\cite{madgraph} and Pythia6~\cite{pythia}, 
and simulated using Delphes~\cite{delphes}.
To construct classifiers for the physics processes, we utilize a total of 23 object-based kinematic variables based on the ATLAS analysis~\cite{atlastth}: 
the transverse momentum (divided by the photon pair invariant mass) and pseudo-rapidity of the two leading photons, 
the transverse momentum, pseudo-rapidity and $b$-tagging status of up to six leading jets, 
as well as the missing transverse momentum.
The signal and background processes differ in the distributions of these variables, providing discriminating power for the machine learning algorithms. Examples of the most powerful variable distributions are shown in Figure~$\ref{fig_training_var}$.
To match the $N$ qubits used by the QSVM-Kernel algorithm, the 23 kinematic variables 
are compressed into \textit{N} variables, using a Principal Component Analysis (PCA) method~\cite{pca,pca1}.
In the case of 15 qubits, for example, each of the 15 variables is formed by combining the 23 original variables.
Afterwards, the \textit{N} variables are rescaled by 
\begin{equation}
    x_i \rightarrow -1 + 2 \times \frac{x_i-x_{i,min}}{x_{i,max}-x_{i,min}}
\end{equation}
where $x_{i,min}$ ($x_{i,max}$) is the minimal (maximal) value of the variable $x_i$, so that the variable values range from $-1$ to $+1$.
This ensures that the rotation angles of $A$, $B$ and $A'$ gates are within $[-1, +1]$, which is found to be slightly more optimal for the \ttH\ analysis than $[-\pi, +\pi]$ used in Ref~\cite{qsvmv}. 

\section{Results}

\textbf{A. Results from Quantum Computer Simulators}

To classify the signal and dominant background processes for the \ttH\ analysis, we employ the QSVM-Kernel algorithm using up to 20 qubits on the \textit{qsim Simulator} from the Google TensorFlow
Quantum framework~\cite{tfq}, the \textit{StatevectorSimulator} from the IBM Quantum framework~\cite{qiskit} and the \textit{Local Simulator} from the Amazon Braket framework~\cite{braket}. 
From this point on, they are referred to as the Google framework, IBM framework and Amazon framework. 
All three simulators model the noiseless execution of quantum computer hardware and evaluate the resulting quantum state vector. 
They represent ideal quantum hardware that can perform infinite measurement shots and experience no hardware device noise.
With training variables processed by PCA, 
we perform our analysis for a number of dataset sizes.
For a given dataset size, we prepare 60 statistically independent datasets to reduce the impact of statistical fluctuations.
Each of the datasets consists of two samples of the same size: a training sample and a test sample. 
(In this study, a dataset of size \textit{n} indicates a training sample of \textit{n} events and a test sample of \textit{n} events.)
We have overcome the challenges of intensive computing resources needed for processing the datasets of up to 50000 events on the quantum computer simulators.
Using the training sample, we adopt a cross-validation procedure~\cite{cv1,cv2} to tune the SVM regularization hyperparameter that controls the size of the margin between the separating hyperplane and the data points in the feature space.  
With the same datasets and the same training variables, we also construct a classical SVM~\cite{svm,svm2} classifier using the scikit-learn package~\cite{sklearn} and a classical BDT~\cite{bdt1,bdt2} classifier using the XGBoost package~\cite{xgboost}. 
The classical SVM and BDT serve as benchmarks for classical machine learning algorithms.
We again perform cross-validation on the training sample to tune the hyper-parameters of the two classical algorithms.
For the classical SVM, we optimize the choice of the classical kernel, the unique hyperparameters for each kernel, and the SVM regularization hyperparameter.
The optimized hyperparameters for the classical BDT include the maximum tree depth and the learning rate.
The other BDT hyperparameters were found to be irrelevant to our study.
\\

\begin{figure}[htb]
\begin{center}
    \includegraphics[width=3.1in]{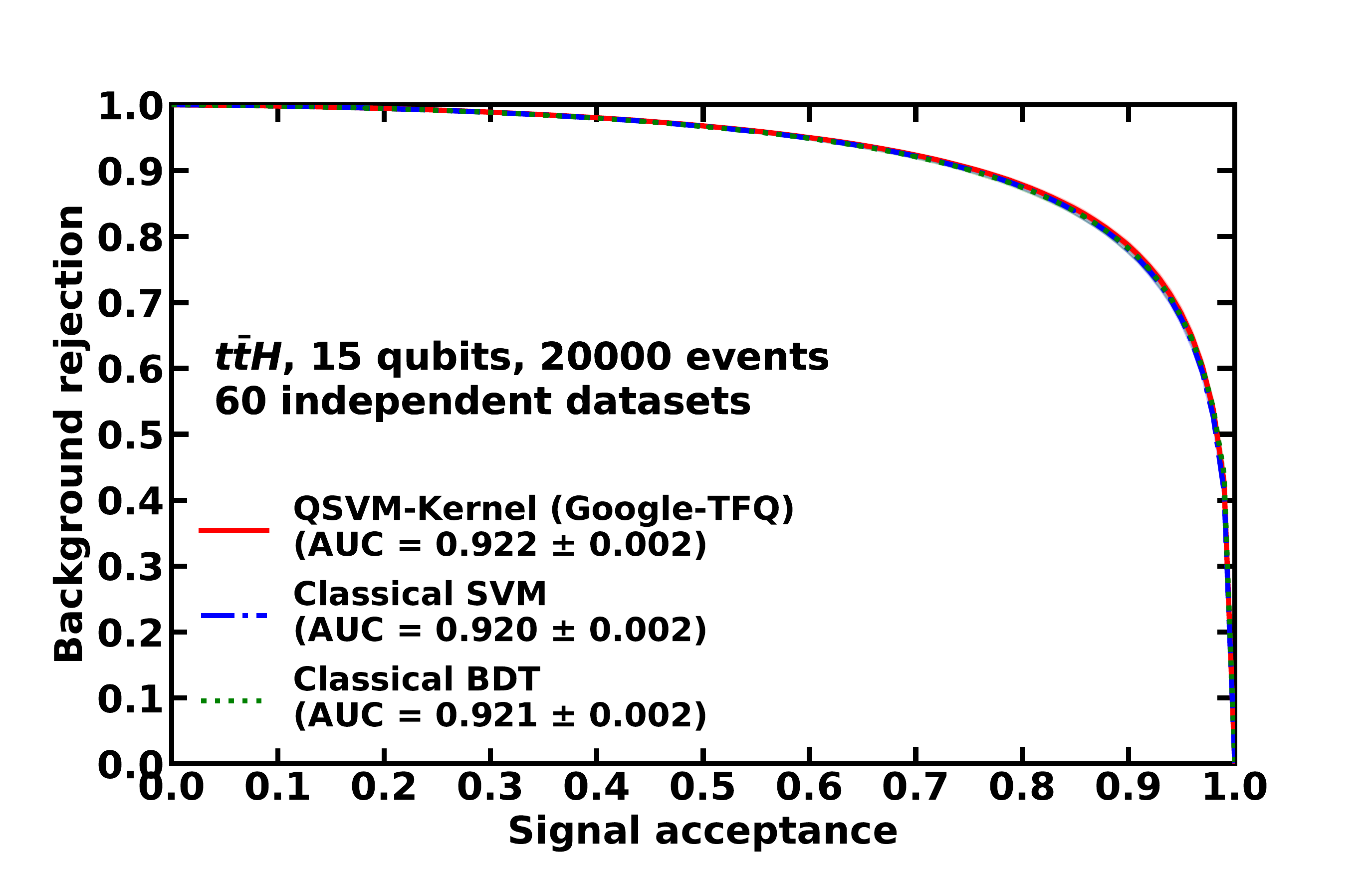} \\
    \textbf{(a)} \\
    \includegraphics[width=3.1in]{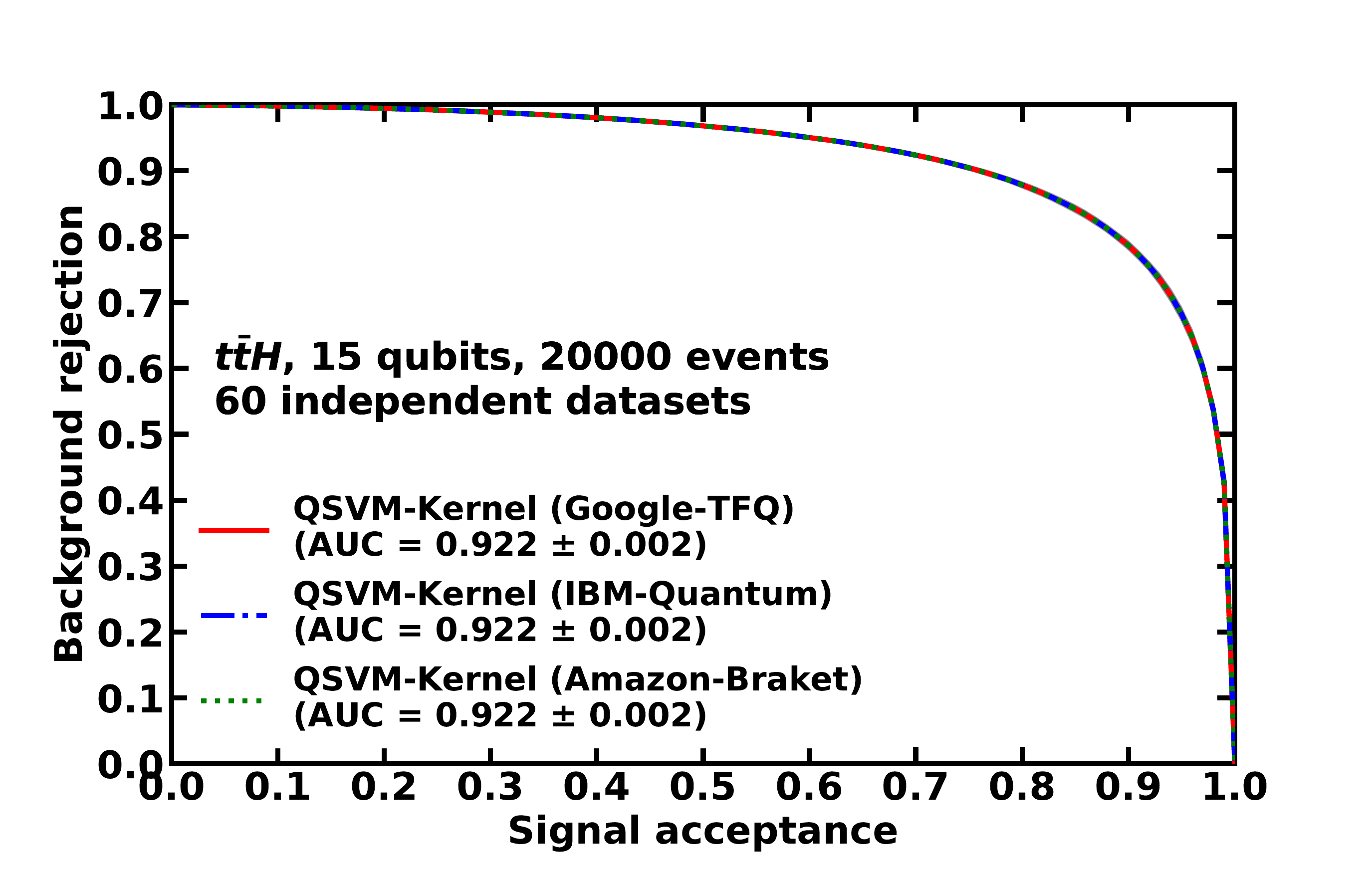} \\
    \textbf{(b)} \\
\end{center}
\caption{
ROC curves of various classifiers using the \ttH\ analysis datasets of 20000 events and 15 input variables.
Each curve represents results averaged over 60 statistically independent datasets.
(a) overlays the results of the QSVM-Kernel algorithm (on the \textit{qsim Simulator} from the Google TensorFlow Quantum framework), the classical SVM algorithm and the classical BDT algorithm.
(b) overlays the QSVM-Kernel results on the \textit{qsim Simulator} from the Google TensorFlow Quantum framework, the \textit{StatevectorSimulator} from the IBM Quantum framework and the \textit{Local Simulator} from the Amazon Braket framework.
Here the QSVM-Kernel classifiers employ 15 qubits on the quantum simulators.
}
\label{fig2}
\end{figure}

To study the discrimination power of each classifier, we produce Receiver Operating Characteristic (ROC) curves that plot background rejection versus signal efficiency,
as well as areas under the ROC curves (AUCs).
ROC curves and AUCs are standard metrics in machine learning applications.
The use of ROC curves and AUCs in this study is inspired by Ref.~\cite{qml4hep0}.
We first show the ROC curves of various classifiers using the \ttH\ analysis datasets of 20000 events and 15 input variables in Figure~\ref{fig2}.
Each curve represents results averaged over 60 statistically independent datasets.
Figure~\ref{fig2} (a) overlays the results of the QSVM-Kernel algorithm (from the Google framework), the classical SVM algorithm and the classical BDT algorithm.
Figure~\ref{fig2} (b) overlays the QSVM-Kernel results from the Google framework, IBM framework and Amazon framework.
Here the QSVM-Kernel classifiers employ 15 qubits on the quantum simulators.
We observe that, 
for these \ttH\ analysis datasets, the QSVM-Kernel performances are similar to the performances given by the two commonly used classical machine learning algorithms.
Furthermore, the three quantum computer simulators, from the Google framework, IBM framework and Amazon framework, provide identical classification performances using the QSVM-Kernel algorithm.\\

Based on the classifier discriminant, we could perform an event selection in order to maximize $S/\sqrt(B)$, where $S$ is the number of selected signal events and $B$ is the number of selected background events. 
$S/\sqrt(B)$ is an approximation of the statistical significance of the signal process, 
and usually correlated with the AUC of the classifier. 
For the above-mentioned QSVM-Kernel classifier in the $\ttH$ analysis, if applying a selection with a signal acceptance of $70\%$, approximately $92\%$ of background events will be rejected and hence $S/\sqrt(B)$ will be improved by around $150\%$ with respect to no selection.
\\

\begin{figure}[htb]
\begin{center}
    \includegraphics[width=2.5in]{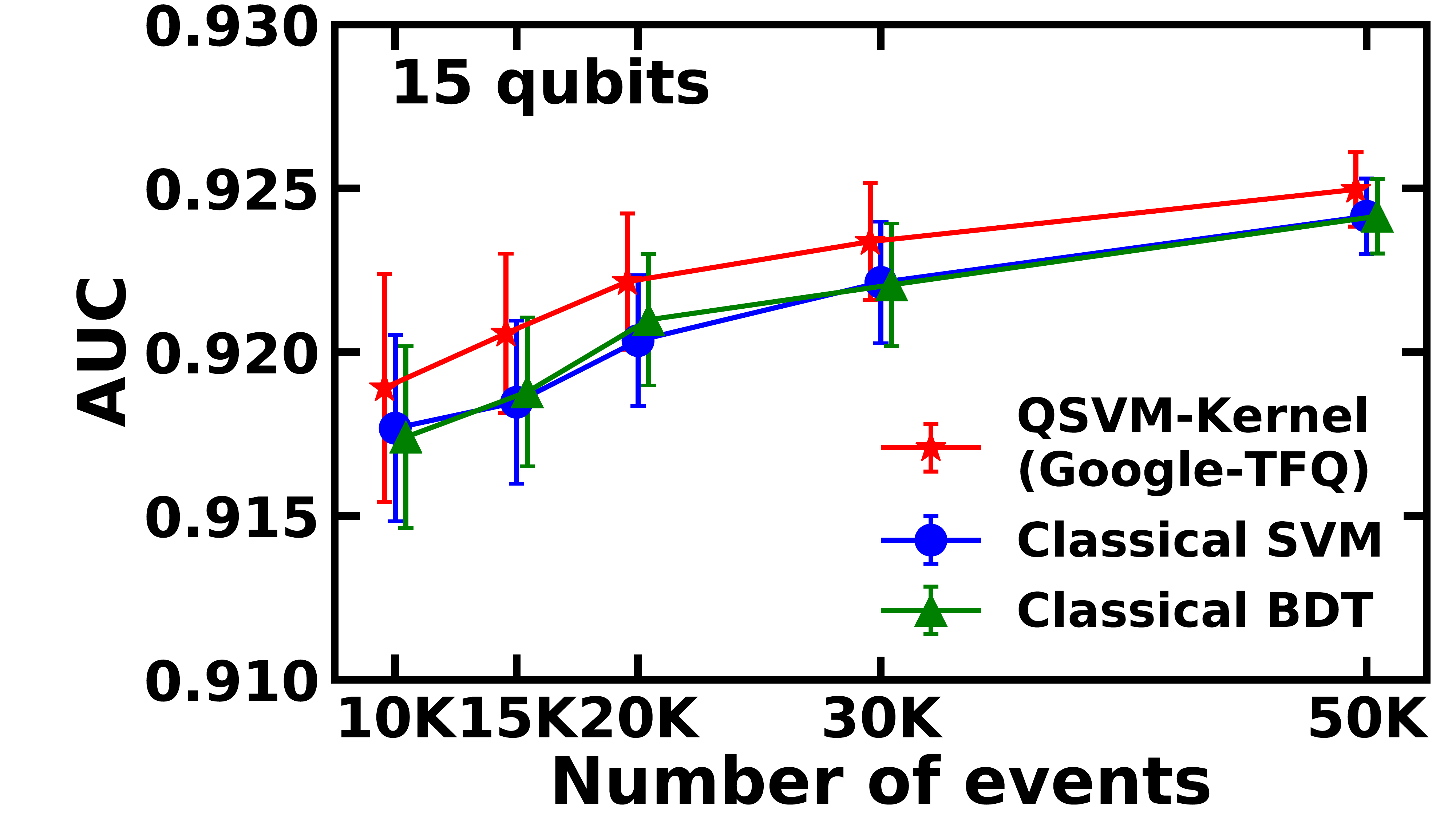} \\
    \textbf{(a)}\\
    \includegraphics[width=2.5in]{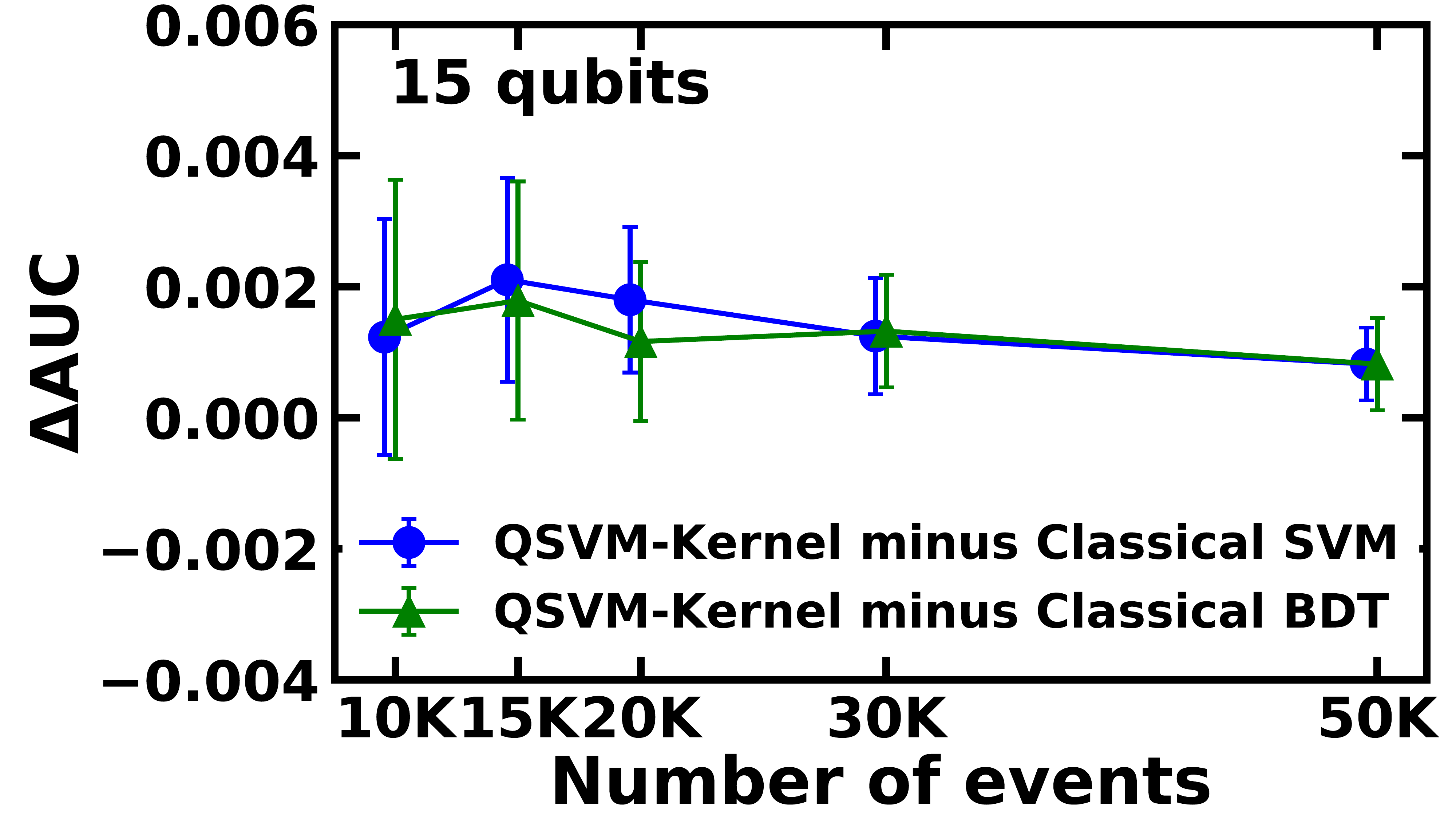} \\
    \textbf{(b)}\\        
    \includegraphics[width=2.5in]{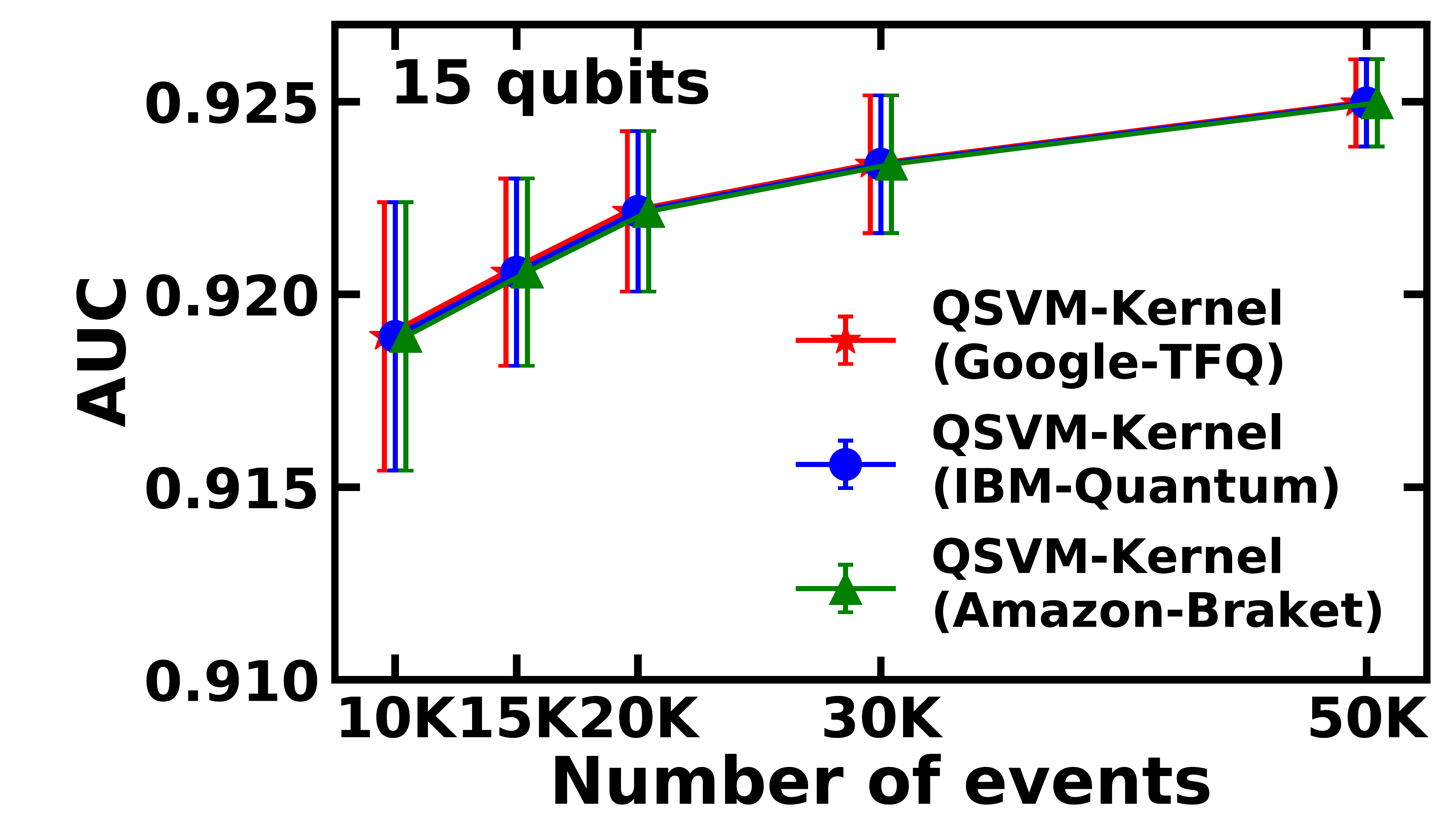} \\ 
    \textbf{(c)}\\
\end{center}
\caption{
The AUC for various classifiers as a function of the \ttH\ analysis dataset size (10000 to 50000 events). 
(a) shows the results of the QSVM-Kernel (on the \textit{qsim Simulator} from the Google TensorFlow Quantum framework), the classical SVM and the classical BDT.
(b) further shows the difference between the QSVM-Kernel algorithm and the classical algorithms.
(c) shows the QSVM-Kernel results on the \textit{qsim Simulator} from the Google TensorFlow Quantum framework, the \textit{StatevectorSimulator} from the IBM Quantum framework and the \textit{Local Simulator} from the Amazon Braket framework.
Here all the classifiers use the same 15 variables and the QSVM-Kernel classifiers employ 15 qubits on the quantum simulators. 
The quoted AUCs are averaged over 60 statistically independent datasets and the quoted errors are the standard deviations for the AUCs of the 60 datasets.
}
\label{fig3}
\end{figure}

Our observation becomes more clear in Figure~\ref{fig3}, where we study the AUC for various classifiers as a function of the \ttH\ analysis dataset size (10000 to 50000 events). 
Figure~\ref{fig3} (a) shows the results of the QSVM-Kernel (from the Google framework), the classical SVM and the classical BDT.
Figure~\ref{fig3} (b) further shows the difference between the QSVM-Kernel algorithm and the classical algorithms.
Figure~\ref{fig3} (c) shows the QSVM-Kernel results from the Google framework, IBM framework and Amazon framework.
Here all the classifiers use the same 15 variables and the QSVM-Kernel classifiers employ 15 qubits on the quantum simulators. 
The quoted AUCs are averaged over 60 statistically independent datasets and the quoted errors are the standard deviations for the AUCs of the 60 datasets.
We find that, the performance of all methods improve with increasing dataset size.
For 15 qubits and up to 50000 events, the QSVM-Kernel algorithm performs similarly to the classical SVM and classical BDT algorithms.
Furthermore, the QSVM-Kernel performances from the three different quantum computer simulators (Google, IBM and Amazon) are comparable.
\\

\begin{figure}[htb]
\begin{center}
    \includegraphics[width=2.5in]{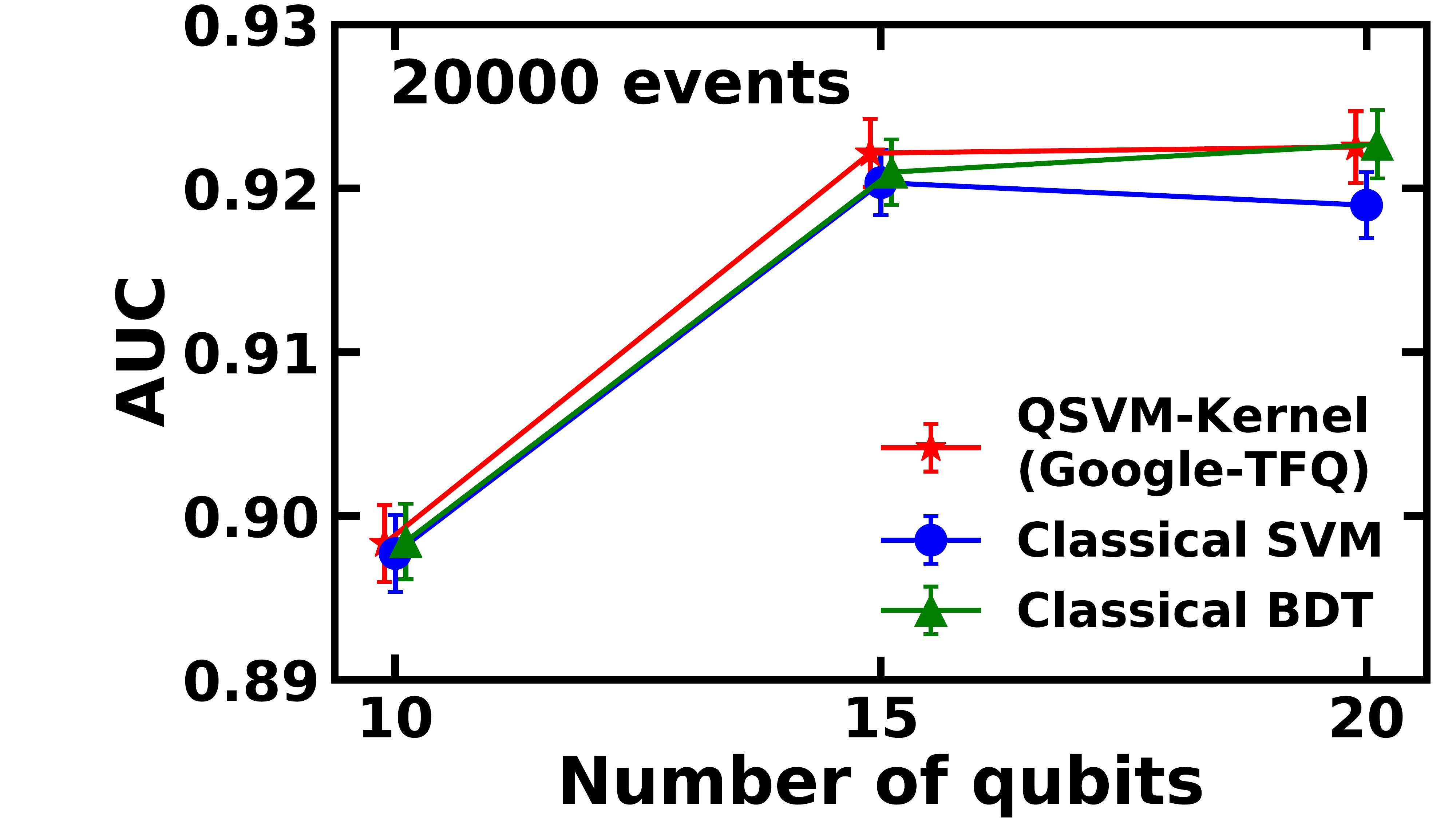}   \\
    \textbf{(a)}\\
    \includegraphics[width=2.5in]{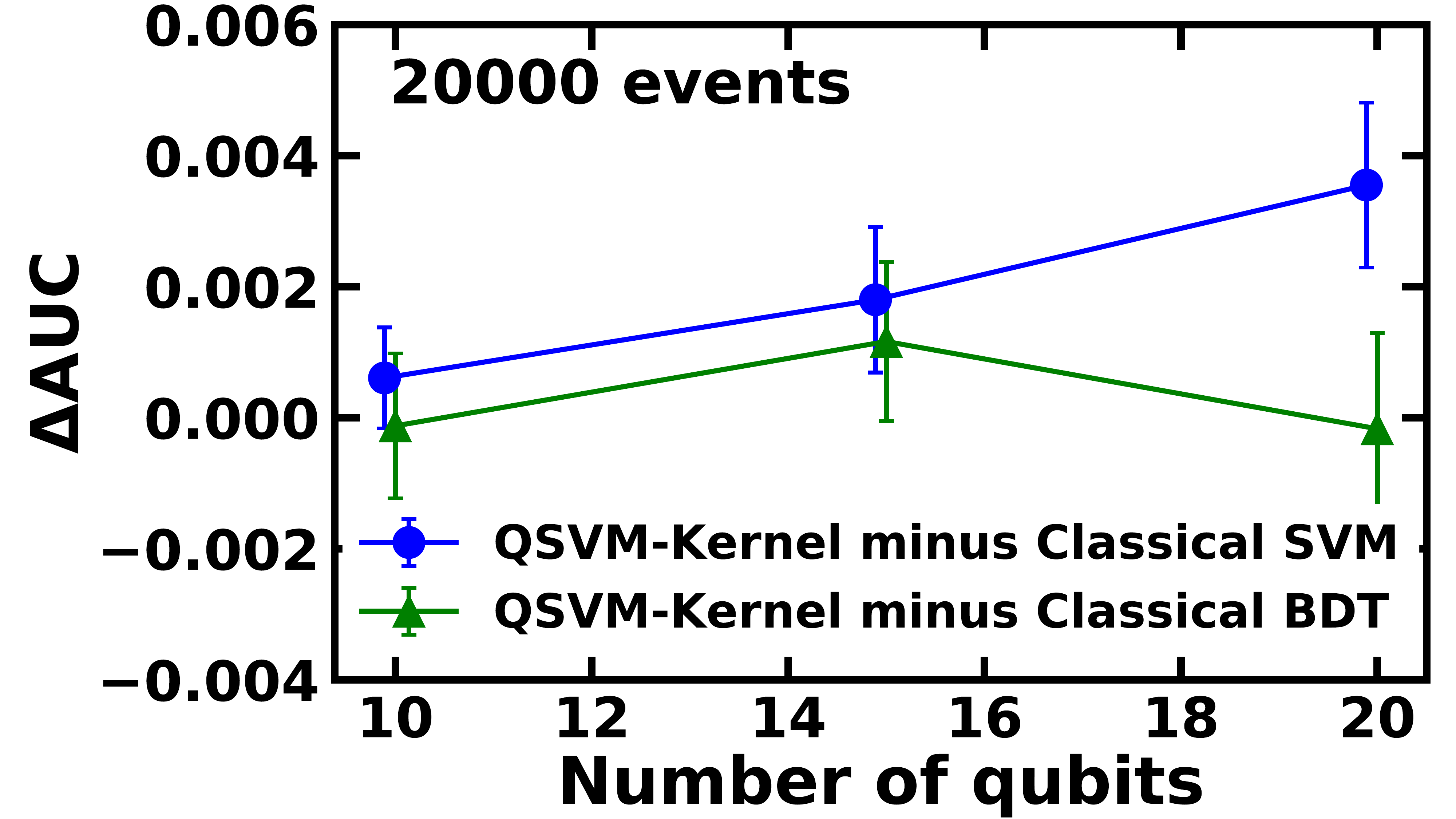} \\
    \textbf{(b)}\\    
    \includegraphics[width=2.5in]{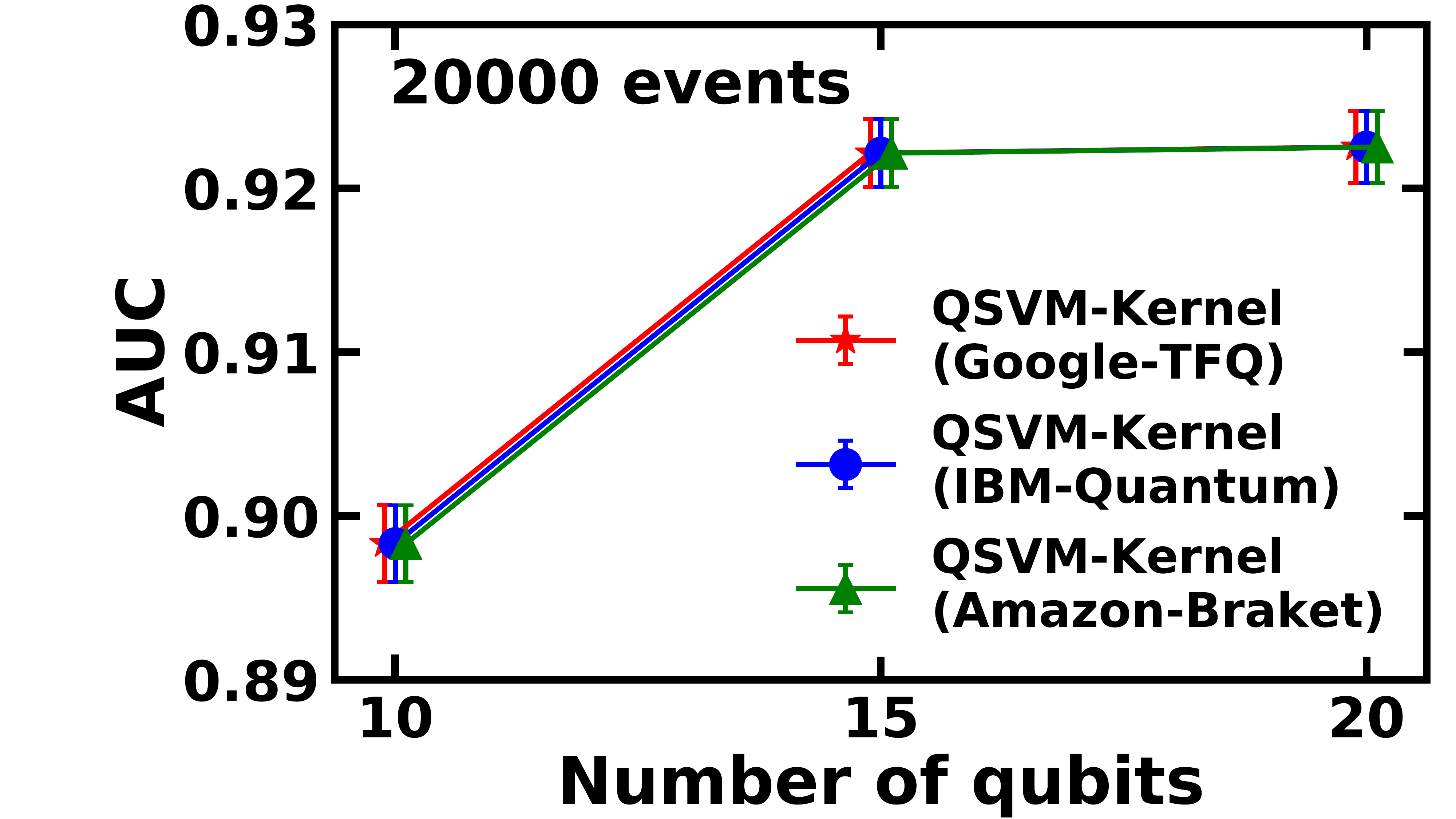} \\
    \textbf{(c)}\\ 
\end{center}
\caption{
AUCs of the QSVM-Kernel algorithm as a function of the number of qubits (10 to 20 qubits).
The number of qubits is equal to the number of input variables.
The 60 statistically independent \ttH\ analysis datasets of 20000 events are used in this study.
In (a), we compare the results of the QSVM-Kernel classifier (on the \textit{qsim Simulator} from the Google TensorFlow Quantum framework) with the results of the classical SVM and classical BDT classifiers using the same input variables. 
In (b), we further display the difference between the QSVM-Kernel results and the classical machine learning results.
In (c), we compare the \textit{qsim Simulator} from the Google TensorFlow Quantum framework, the \textit{StatevectorSimulator} from the IBM Quantum framework and the \textit{Local Simulator} from the Amazon Braket framework for the QSVM-Kernel results.
}
\label{fig_auc_vs_nvars}
\end{figure}

We also investigate the AUCs of the QSVM-Kernel algorithm as a function of the number of qubits (10 to 20 qubits), as shown in Figure~$\ref{fig_auc_vs_nvars}$.
The number of qubits is equal to the number of input variables using PCA as described in Section 3.
The 60 statistically independent \ttH\ analysis datasets of 20000 events are used in this study.
In Figure~$\ref{fig_auc_vs_nvars}$ (a), we compare the results of the QSVM-Kernel (from the Google framework) with the classical SVM and classical BDT using the same input variables. 
In Figure~$\ref{fig_auc_vs_nvars}$ (b), we further display the difference between the QSVM-Kernel results and the classical machine learning results.
In Figure~$\ref{fig_auc_vs_nvars}$ (c), again, we compare the Google framework, IBM framework and Amazon framework for the QSVM-Kernel results.
We find that, the QSVM-Kernel result with 15 qubits is better than 10 qubits and similar to 20 qubits.
For 10 to 20 qubits and 20000 events, the performance of the QSVM-Kernel algorithm is similar to that of the classical SVM algorithm. 
Again, the three quantum computer simulators (Google, IBM and Amazon) yield the same classification power.
\\

\textbf{B. Results from Quantum Computer Hardware}

\begin{figure}[htb]
\begin{center}
\includegraphics[width=3.1in]{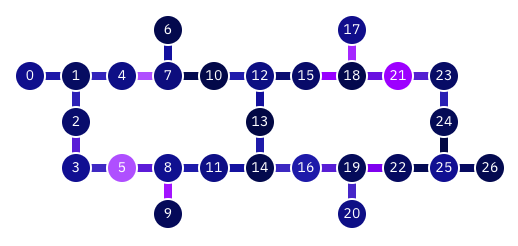}\\
\end{center}
\caption{
The qubit map of the \textit{``ibmq\_paris''} quantum system~\cite{ibm_web}. The (darker) colors indicate (lower) readout error rates of the qubits and CNOT error rates of the connections. Our study uses qubits 3, 5, 8, 11, 14, 16, 19, 22, 25, 24, 23, 21, 18, 15 and 12.
}
\label{fig_paris}
\end{figure}

After the studies using simulation of the ideal quantum computers, it is now of great interest to assess the quantum machine learning performances on today's noisy quantum computer hardware. 
For the \ttH\ physics analysis, 
we employ the QSVM-Kernel algorithm 
on the IBM \textit{``ibmq\_paris''} quantum computer hardware. 
\textit{``ibmq\_paris''} is a 27-qubit quantum processor based on superconducting electronic circuits. 
The qubit map of the \textit{``ibmq\_paris''} quantum system~\cite{ibm_web} is shown in Figure~\ref{fig_paris}. 
Due to limited access time available to us, we performed six runs using 15 qubits on \textit{``ibmq\_paris''}.
Each run processes a statistically independent dataset of 100 events.
For these six runs, the average running time on the quantum hardware is approximately 680 minutes.
With more advanced quantum hardware in the future, the running time is expected to be significantly reduced.
The quantum circuit of the hardware runs is kept the same as for the simulator runs, 
while the SVM regularization hyperparameter is separately optimized for hardware and simulator runs.
To reduce statistical uncertainties in evaluating kernel entries on quantum hardware, we use 8192 measurement shots for every kernel entry. 
\\

\begin{figure}[htb]
\begin{center}
\includegraphics[width=3.0in]{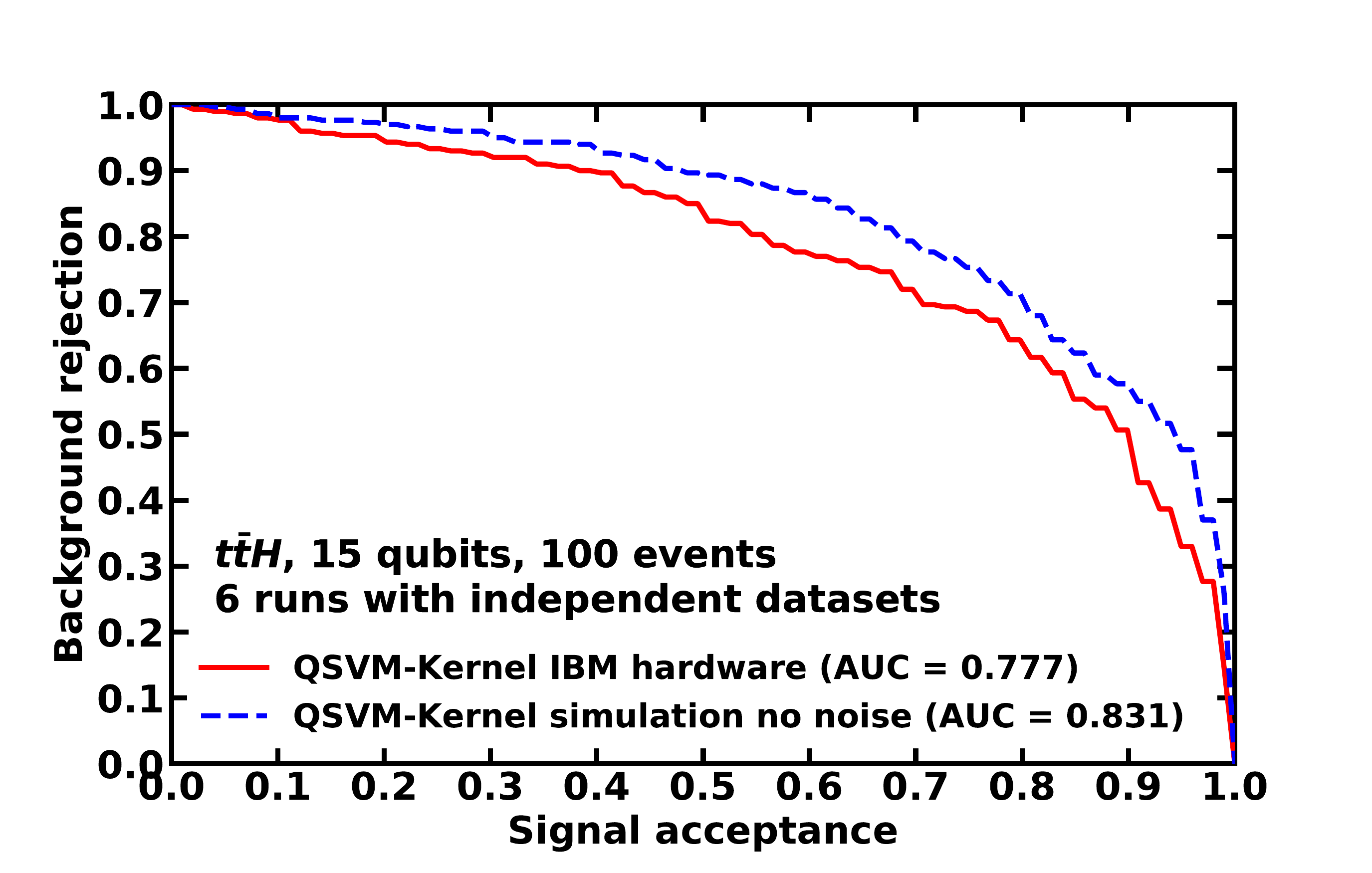}\\
\end{center}
\caption{
ROC curve of the QSVM-Kernel classifier with the \textit{``ibmq\_paris''} quantum computer hardware using the \ttH\ analysis datasets of 100 events.
For comparison, we overlay the ROC curve with the \textit{StatevectorSimulator} from the IBM Quantum framework using the same datasets.
The results are averaged over the six hardware runs.
All the QSVM-Kernel classifiers use 15 qubits and the same 15 variables. 
}
\label{fig4}
\end{figure}

\begin{figure}[htb]
\begin{center}
\includegraphics[width=3.0in]{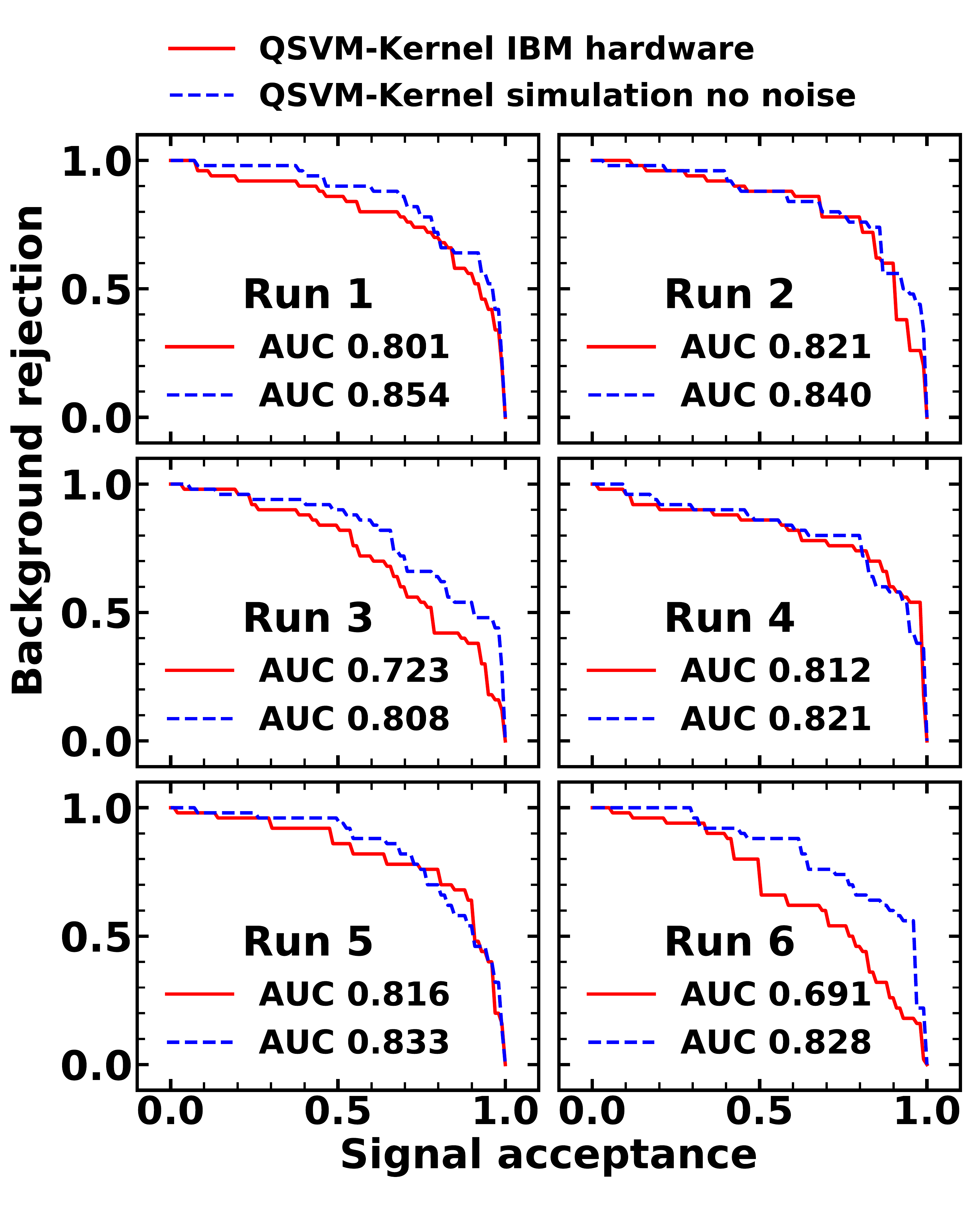}\\
\end{center}
\caption{
ROC curve with the \textit{``ibmq\_paris''} quantum computer hardware and ROC curve with the \textit{StatevectorSimulator} from the IBM Quantum framework for each of the six hardware runs.
Each run processes a statistically independent dataset of 100 events.
All the QSVM-Kernel classifiers are using 15 qubits and the same 15 variables. 
}
\label{fig5}
\end{figure}

In Figure~\ref{fig4}, we present the ROC curve of the QSVM-Kernel classifier with the \textit{``ibmq\_paris''} quantum computer hardware using the \ttH\ analysis datasets of 100 events.
For comparison, we overlay the ROC curve with the \textit{StatevectorSimulator} from the IBM Quantum framework using the same datasets.
The results are averaged over the six hardware runs.
All the QSVM-Kernel classifiers use 15 qubits and the same 15 variables. 
In Figure~\ref{fig5}, we compare the ROC curve with the \textit{``ibmq\_paris''} quantum computer hardware and the ROC curve with the \textit{StatevectorSimulator} for each of the six hardware runs.
With small training samples of 100 events,  
the performance achieved by the \textit{``ibmq\_paris''} quantum computer hardware is promising and approaching the noiseless quantum computer simulator.
The difference between the hardware performance and the simulator performance is likely due to the effect of quantum hardware noise and fluctuates among our hardware runs.

\section{Conclusion}

In this study, we have successfully employed the quantum support vector machine kernel (QSVM-Kernel) method in the \ttH\ (Higgs boson production in association with a top quark pair) physics analysis, a recent LHC flagship physics analysis, 
on gate-model quantum computer simulators and hardware. 
The simulation study has been performed using the Google TensorFlow Quantum framework, IBM Quantum framework and Amazon Braket framework.
We have overcome the challenges of intensive computing resources in the cases of up to 20 qubits and up to 50000 events on the quantum computer simulators, in order to perform quantum machine learning studies on physics datasets that closely resemble those used in the official ATLAS publication~\cite{atlastth}. 
The QSVM-Kernel method achieves good classification performance that is similar to the performances of the classical machine learning methods currently used in LHC physics analyses, classical SVM and classical BDT for example.
On the \textit{``ibmq\_paris''} superconducting quantum computer hardware, we have also employed the QSVM-Kernel algorithm using 100 events and 15 qubits to assess the effect of quantum hardware noise. 
The performance achieved on the \textit{``ibmq\_paris''} quantum hardware is promising and is approaching the performance from the noiseless quantum simulators. 

Our quantum simulation result gives an example that quantum machine learning performs as well as its classical counterpart using three different platforms (Google, IBM and Amazon) 
for realistic high energy physics analysis datasets. 
Furthermore, our result on noisy quantum hardware provides important validation for the result on noiseless quantum simulators.
Our studies confirm that the QSVM-Kernel algorithm can use the large dimensionality of the quantum Hilbert space to replace the classical feature space. 
In the future, large improvement in computational speed and reduction in device noise on quantum computing hardware will likely be achieved and lead to quantum advantage in quantum machine learning applications. With the large investments in quantum computing and fierce competitions in technology, this expectation is realistic. Therefore, we predict that quantum machine learning will become a powerful tool for data analysis in High Energy Physics.


\begin{thebibliography}{40}%
\makeatletter
\providecommand \@ifxundefined [1]{%
 \@ifx{#1\undefined}
}%
\providecommand \@ifnum [1]{%
 \ifnum #1\expandafter \@firstoftwo
 \else \expandafter \@secondoftwo
 \fi
}%
\providecommand \@ifx [1]{%
 \ifx #1\expandafter \@firstoftwo
 \else \expandafter \@secondoftwo
 \fi
}%
\providecommand \natexlab [1]{#1}%
\providecommand \enquote  [1]{``#1''}%
\providecommand \bibnamefont  [1]{#1}%
\providecommand \bibfnamefont [1]{#1}%
\providecommand \citenamefont [1]{#1}%
\providecommand \href@noop [0]{\@secondoftwo}%
\providecommand \href [0]{\begingroup \@sanitize@url \@href}%
\providecommand \@href[1]{\@@startlink{#1}\@@href}%
\providecommand \@@href[1]{\endgroup#1\@@endlink}%
\providecommand \@sanitize@url [0]{\catcode `\\12\catcode `\$12\catcode
  `\&12\catcode `\#12\catcode `\^12\catcode `\_12\catcode `\%12\relax}%
\providecommand \@@startlink[1]{}%
\providecommand \@@endlink[0]{}%
\providecommand \url  [0]{\begingroup\@sanitize@url \@url }%
\providecommand \@url [1]{\endgroup\@href {#1}{\urlprefix }}%
\providecommand \urlprefix  [0]{URL }%
\providecommand \Eprint [0]{\href }%
\providecommand \doibase [0]{https://doi.org/}%
\providecommand \selectlanguage [0]{\@gobble}%
\providecommand \bibinfo  [0]{\@secondoftwo}%
\providecommand \bibfield  [0]{\@secondoftwo}%
\providecommand \translation [1]{[#1]}%
\providecommand \BibitemOpen [0]{}%
\providecommand \bibitemStop [0]{}%
\providecommand \bibitemNoStop [0]{.\EOS\space}%
\providecommand \EOS [0]{\spacefactor3000\relax}%
\providecommand \BibitemShut  [1]{\csname bibitem#1\endcsname}%
\let\auto@bib@innerbib\@empty
\bibitem [{\citenamefont {{ATLAS Collaboration G. Aad et
  al.}}(2012)}]{atlashiggs}%
  \BibitemOpen
  \bibfield  {author} {\bibinfo {author} {\bibnamefont {{ATLAS Collaboration G.
  Aad et al.}}},\ }\bibfield  {title} {\bibinfo {title} {{Observation of a new
  particle in the search for the standard model Higgs boson with the ATLAS
  detector at the LHC}},\ }\href@noop {} {\bibfield  {journal} {\bibinfo
  {journal} {Physics Letters B}\ }\textbf {\bibinfo {volume} {716}},\ \bibinfo
  {pages} {1} (\bibinfo {year} {2012})}\BibitemShut {NoStop}%
\bibitem [{\citenamefont {{CMS Collaboration, S. Chatrchyan et
  al.}}(2012)}]{cmshiggs}%
  \BibitemOpen
  \bibfield  {author} {\bibinfo {author} {\bibnamefont {{CMS Collaboration, S.
  Chatrchyan et al.}}},\ }\bibfield  {title} {\bibinfo {title} {{Observation of
  a new boson at a mass of 125 GeV with the CMS experiment at the LHC}},\
  }\href@noop {} {\bibfield  {journal} {\bibinfo  {journal} {Physics Letters
  B}\ }\textbf {\bibinfo {volume} {716}},\ \bibinfo {pages} {30} (\bibinfo
  {year} {2012})}\BibitemShut {NoStop}%
\bibitem [{\citenamefont {Radovic}\ \emph {et~al.}(2018)\citenamefont
  {Radovic}, \citenamefont {Williams}, \citenamefont {Rousseau}, \citenamefont
  {Kagan}, \citenamefont {Bonacorsi}, \citenamefont {Himmel}, \citenamefont
  {Aurisano}, \citenamefont {Terao},\ and\ \citenamefont
  {Wongjirad}}]{ml4hep1}%
  \BibitemOpen
  \bibfield  {author} {\bibinfo {author} {\bibfnamefont {A.}~\bibnamefont
  {Radovic}}, \bibinfo {author} {\bibfnamefont {M.}~\bibnamefont {Williams}},
  \bibinfo {author} {\bibfnamefont {D.}~\bibnamefont {Rousseau}}, \bibinfo
  {author} {\bibfnamefont {M.}~\bibnamefont {Kagan}}, \bibinfo {author}
  {\bibfnamefont {D.}~\bibnamefont {Bonacorsi}}, \bibinfo {author}
  {\bibfnamefont {A.}~\bibnamefont {Himmel}}, \bibinfo {author} {\bibfnamefont
  {A.}~\bibnamefont {Aurisano}}, \bibinfo {author} {\bibfnamefont
  {K.}~\bibnamefont {Terao}},\ and\ \bibinfo {author} {\bibfnamefont
  {T.}~\bibnamefont {Wongjirad}},\ }\bibfield  {title} {\bibinfo {title}
  {{Machine learning at the energy and intensity frontiers of particle
  physics}},\ }\href@noop {} {\bibfield  {journal} {\bibinfo  {journal}
  {Nature}\ }\textbf {\bibinfo {volume} {560}},\ \bibinfo {pages} {41}
  (\bibinfo {year} {2018})}\BibitemShut {NoStop}%
\bibitem [{\citenamefont {Albertsson}\ \emph {et~al.}(2018)\citenamefont
  {Albertsson} \emph {et~al.}}]{ml4hep2}%
  \BibitemOpen
  \bibfield  {author} {\bibinfo {author} {\bibfnamefont {K.}~\bibnamefont
  {Albertsson}} \emph {et~al.},\ }\bibfield  {title} {\bibinfo {title}
  {{Machine Learning in High Energy Physics Community White Paper}},\ }\href
  {https://doi.org/10.1088/1742-6596/1085/2/022008} {\bibfield  {journal}
  {\bibinfo  {journal} {J. Phys. Conf. Ser.}\ }\textbf {\bibinfo {volume}
  {1085}},\ \bibinfo {pages} {022008} (\bibinfo {year} {2018})},\ \Eprint
  {https://arxiv.org/abs/1807.02876} {arXiv:1807.02876 [physics.comp-ph]}
  \BibitemShut {NoStop}%
\bibitem [{\citenamefont {Guest}\ \emph {et~al.}(2018)\citenamefont {Guest},
  \citenamefont {Cranmer},\ and\ \citenamefont {Whiteson}}]{ml4hep3}%
  \BibitemOpen
  \bibfield  {author} {\bibinfo {author} {\bibfnamefont {D.}~\bibnamefont
  {Guest}}, \bibinfo {author} {\bibfnamefont {K.}~\bibnamefont {Cranmer}},\
  and\ \bibinfo {author} {\bibfnamefont {D.}~\bibnamefont {Whiteson}},\
  }\bibfield  {title} {\bibinfo {title} {{Deep Learning and its Application to
  LHC Physics}},\ }\href {https://doi.org/10.1146/annurev-nucl-101917-021019}
  {\bibfield  {journal} {\bibinfo  {journal} {Ann. Rev. Nucl. Part. Sci.}\
  }\textbf {\bibinfo {volume} {68}},\ \bibinfo {pages} {161} (\bibinfo {year}
  {2018})},\ \Eprint {https://arxiv.org/abs/1806.11484} {arXiv:1806.11484
  [hep-ex]} \BibitemShut {NoStop}%
\bibitem [{\citenamefont {Carleo}\ \emph {et~al.}(2019)\citenamefont {Carleo},
  \citenamefont {Cirac}, \citenamefont {Cranmer}, \citenamefont {Daudet},
  \citenamefont {Schuld}, \citenamefont {Tishby}, \citenamefont
  {Vogt-Maranto},\ and\ \citenamefont {Zdeborov\'a}}]{ml4hep4}%
  \BibitemOpen
  \bibfield  {author} {\bibinfo {author} {\bibfnamefont {G.}~\bibnamefont
  {Carleo}}, \bibinfo {author} {\bibfnamefont {I.}~\bibnamefont {Cirac}},
  \bibinfo {author} {\bibfnamefont {K.}~\bibnamefont {Cranmer}}, \bibinfo
  {author} {\bibfnamefont {L.}~\bibnamefont {Daudet}}, \bibinfo {author}
  {\bibfnamefont {M.}~\bibnamefont {Schuld}}, \bibinfo {author} {\bibfnamefont
  {N.}~\bibnamefont {Tishby}}, \bibinfo {author} {\bibfnamefont
  {L.}~\bibnamefont {Vogt-Maranto}},\ and\ \bibinfo {author} {\bibfnamefont
  {L.}~\bibnamefont {Zdeborov\'a}},\ }\bibfield  {title} {\bibinfo {title}
  {{Machine learning and the physical sciences}},\ }\href
  {https://doi.org/10.1103/RevModPhys.91.045002} {\bibfield  {journal}
  {\bibinfo  {journal} {Rev. Mod. Phys.}\ }\textbf {\bibinfo {volume} {91}},\
  \bibinfo {pages} {045002} (\bibinfo {year} {2019})},\ \Eprint
  {https://arxiv.org/abs/1903.10563} {arXiv:1903.10563 [physics.comp-ph]}
  \BibitemShut {NoStop}%
\bibitem [{\citenamefont {Bourilkov}(2020)}]{ml4hep5}%
  \BibitemOpen
  \bibfield  {author} {\bibinfo {author} {\bibfnamefont {D.}~\bibnamefont
  {Bourilkov}},\ }\bibfield  {title} {\bibinfo {title} {{Machine and Deep
  Learning Applications in Particle Physics}},\ }\href
  {https://doi.org/10.1142/S0217751X19300199} {\bibfield  {journal} {\bibinfo
  {journal} {Int. J. Mod. Phys. A}\ }\textbf {\bibinfo {volume} {34}},\
  \bibinfo {pages} {1930019} (\bibinfo {year} {2020})},\ \Eprint
  {https://arxiv.org/abs/1912.08245} {arXiv:1912.08245 [physics.data-an]}
  \BibitemShut {NoStop}%
\bibitem [{\citenamefont {{ATLAS Collaboration, M. Aaboud et
  al.}}(2018)}]{atlastth}%
  \BibitemOpen
  \bibfield  {author} {\bibinfo {author} {\bibnamefont {{ATLAS Collaboration,
  M. Aaboud et al.}}},\ }\bibfield  {title} {\bibinfo {title} {{Observation of
  Higgs boson production in association with a top quark pair at the LHC with
  the ATLAS detector}},\ }\href@noop {} {\bibfield  {journal} {\bibinfo
  {journal} {Physics Letters B}\ }\textbf {\bibinfo {volume} {784}},\ \bibinfo
  {pages} {173} (\bibinfo {year} {2018})}\BibitemShut {NoStop}%
\bibitem [{\citenamefont {{CMS Collaboration, A. M. Sirunyan et
  al.}}(2018)}]{cmstth}%
  \BibitemOpen
  \bibfield  {author} {\bibinfo {author} {\bibnamefont {{CMS Collaboration, A.
  M. Sirunyan et al.}}},\ }\bibfield  {title} {\bibinfo {title} {{Observation
  of $t\overline{t}H$ production}},\ }\href@noop {} {\bibfield  {journal}
  {\bibinfo  {journal} {Physical Review Letters}\ }\textbf {\bibinfo {volume}
  {120}},\ \bibinfo {pages} {231801} (\bibinfo {year} {2018})}\BibitemShut
  {NoStop}%
\bibitem [{\citenamefont {Biamonte}\ \emph {et~al.}(2017)\citenamefont
  {Biamonte}, \citenamefont {Wittek}, \citenamefont {Pancotti}, \citenamefont
  {Rebentrost}, \citenamefont {Wiebe},\ and\ \citenamefont {Lloyd}}]{qml}%
  \BibitemOpen
  \bibfield  {author} {\bibinfo {author} {\bibfnamefont {J.}~\bibnamefont
  {Biamonte}}, \bibinfo {author} {\bibfnamefont {P.}~\bibnamefont {Wittek}},
  \bibinfo {author} {\bibfnamefont {N.}~\bibnamefont {Pancotti}}, \bibinfo
  {author} {\bibfnamefont {P.}~\bibnamefont {Rebentrost}}, \bibinfo {author}
  {\bibfnamefont {N.}~\bibnamefont {Wiebe}},\ and\ \bibinfo {author}
  {\bibfnamefont {S.}~\bibnamefont {Lloyd}},\ }\bibfield  {title} {\bibinfo
  {title} {{Quantum machine learning}},\ }\href@noop {} {\bibfield  {journal}
  {\bibinfo  {journal} {Nature}\ }\textbf {\bibinfo {volume} {549}},\ \bibinfo
  {pages} {195–202} (\bibinfo {year} {2017})}\BibitemShut {NoStop}%
\bibitem [{\citenamefont {Huang}\ \emph {et~al.}(2021)\citenamefont {Huang},
  \citenamefont {Broughton}, \citenamefont {Mohseni}, \citenamefont {Babbush},
  \citenamefont {Boixo}, \citenamefont {Neven},\ and\ \citenamefont
  {McClean}}]{qml-pa}%
  \BibitemOpen
  \bibfield  {author} {\bibinfo {author} {\bibfnamefont {H.-Y.}\ \bibnamefont
  {Huang}}, \bibinfo {author} {\bibfnamefont {M.}~\bibnamefont {Broughton}},
  \bibinfo {author} {\bibfnamefont {M.}~\bibnamefont {Mohseni}}, \bibinfo
  {author} {\bibfnamefont {R.}~\bibnamefont {Babbush}}, \bibinfo {author}
  {\bibfnamefont {S.}~\bibnamefont {Boixo}}, \bibinfo {author} {\bibfnamefont
  {H.}~\bibnamefont {Neven}},\ and\ \bibinfo {author} {\bibfnamefont {J.~R.}\
  \bibnamefont {McClean}},\ }\bibfield  {title} {\bibinfo {title} {Power of
  data in quantum machine learning},\ }\href@noop {} {\bibfield  {journal}
  {\bibinfo  {journal} {Nature Communications}\ }\textbf {\bibinfo {volume}
  {12}},\ \bibinfo {pages} {2631} (\bibinfo {year} {2021})}\BibitemShut
  {NoStop}%
\bibitem [{\citenamefont {Guan}\ \emph {et~al.}(2020)\citenamefont {Guan},
  \citenamefont {Perdue}, \citenamefont {Pesah}, \citenamefont {Schuld},
  \citenamefont {Terashi}, \citenamefont {Vallecorsa},\ and\ \citenamefont
  {Vlimant}}]{qml-hep}%
  \BibitemOpen
  \bibfield  {author} {\bibinfo {author} {\bibfnamefont {W.}~\bibnamefont
  {Guan}}, \bibinfo {author} {\bibfnamefont {G.}~\bibnamefont {Perdue}},
  \bibinfo {author} {\bibfnamefont {A.}~\bibnamefont {Pesah}}, \bibinfo
  {author} {\bibfnamefont {M.}~\bibnamefont {Schuld}}, \bibinfo {author}
  {\bibfnamefont {K.}~\bibnamefont {Terashi}}, \bibinfo {author} {\bibfnamefont
  {S.}~\bibnamefont {Vallecorsa}},\ and\ \bibinfo {author} {\bibfnamefont
  {J.-R.}\ \bibnamefont {Vlimant}},\ }\bibfield  {title} {\bibinfo {title}
  {{Quantum Machine Learning in High Energy Physics}},\ }\href@noop {} {\
  (\bibinfo {year} {2020})},\ \Eprint {https://arxiv.org/abs/2005.08582}
  {arXiv:2005.08582 [quant-ph]} \BibitemShut {NoStop}%
\bibitem [{\citenamefont {Gambetta}(2020)}]{ibm-roadmap}%
  \BibitemOpen
  \bibfield  {author} {\bibinfo {author} {\bibfnamefont {J.}~\bibnamefont
  {Gambetta}},\ }\href@noop {} {\bibinfo {title} {Ibm’s roadmap for scaling
  quantum technology}} (\bibinfo {year} {2020}),\ \bibinfo {note}
  {\\https://www.ibm.com/blogs/research/2020/09/ibm-quantum-roadmap/}\BibitemShut
  {NoStop}%
\bibitem [{\citenamefont {Shankland}(2019)}]{google-roadmap}%
  \BibitemOpen
  \bibfield  {author} {\bibinfo {author} {\bibfnamefont {S.}~\bibnamefont
  {Shankland}},\ }\href@noop {} {\bibinfo {title} {Quantum computer makers like
  their odds for big progress}} (\bibinfo {year} {2019}),\ \bibinfo {note}
  {\\https://www.cnet.com/tech/computing/quantum-computer-makers-like-their-odds-for-big-progress-soon/}\BibitemShut
  {NoStop}%
\bibitem [{\citenamefont {Hubs}()}]{ionq-roadmap}%
  \BibitemOpen
  \bibfield  {author} {\bibinfo {author} {\bibfnamefont {T.~Q.}\ \bibnamefont
  {Hubs}},\ }\href@noop {} {\bibinfo {title} {Ionq’s roadmap up to 2025}},\
  \bibinfo {note}
  {\\https://thequantumhubs.com/ionqs-roadmap-up-to-2025-2/}\BibitemShut
  {NoStop}%
\bibitem [{Apo(2017)}]{Apollinari-2017lan}%
  \BibitemOpen
  \bibfield  {title} {\bibinfo {title} {{High-Luminosity Large Hadron Collider
  (HL-LHC)}: {Technical Design Report V. 0.1}}\ }\textbf {\bibinfo {volume}
  {4/2017}},\ \href {https://doi.org/10.23731/CYRM-2017-004}
  {10.23731/CYRM-2017-004} (\bibinfo {year} {2017})\BibitemShut {NoStop}%
\bibitem [{\citenamefont {Mott}\ \emph {et~al.}(2017)\citenamefont {Mott},
  \citenamefont {Job}, \citenamefont {Vlimant}, \citenamefont {Lidar},\ and\
  \citenamefont {Spiropulu}}]{qml4hep0}%
  \BibitemOpen
  \bibfield  {author} {\bibinfo {author} {\bibfnamefont {A.}~\bibnamefont
  {Mott}}, \bibinfo {author} {\bibfnamefont {J.}~\bibnamefont {Job}}, \bibinfo
  {author} {\bibfnamefont {J.-R.}\ \bibnamefont {Vlimant}}, \bibinfo {author}
  {\bibfnamefont {D.}~\bibnamefont {Lidar}},\ and\ \bibinfo {author}
  {\bibfnamefont {M.}~\bibnamefont {Spiropulu}},\ }\bibfield  {title} {\bibinfo
  {title} {{Solving a Higgs optimization problem with quantum annealing for
  machine learning}},\ }\href@noop {} {\bibfield  {journal} {\bibinfo
  {journal} {Nature}\ }\textbf {\bibinfo {volume} {550}},\ \bibinfo {pages}
  {375} (\bibinfo {year} {2017})}\BibitemShut {NoStop}%
\bibitem [{\citenamefont {Terashi}\ \emph {et~al.}(2021)\citenamefont
  {Terashi}, \citenamefont {Kaneda}, \citenamefont {Kishimoto}, \citenamefont
  {Saito}, \citenamefont {Sawada},\ and\ \citenamefont {Tanaka}}]{qml4hep1}%
  \BibitemOpen
  \bibfield  {author} {\bibinfo {author} {\bibfnamefont {K.}~\bibnamefont
  {Terashi}}, \bibinfo {author} {\bibfnamefont {M.}~\bibnamefont {Kaneda}},
  \bibinfo {author} {\bibfnamefont {T.}~\bibnamefont {Kishimoto}}, \bibinfo
  {author} {\bibfnamefont {M.}~\bibnamefont {Saito}}, \bibinfo {author}
  {\bibfnamefont {R.}~\bibnamefont {Sawada}},\ and\ \bibinfo {author}
  {\bibfnamefont {J.}~\bibnamefont {Tanaka}},\ }\bibfield  {title} {\bibinfo
  {title} {{Event Classification with Quantum Machine Learning in High-Energy
  Physics}},\ }\href {https://doi.org/10.1007/s41781-020-00047-7} {\bibfield
  {journal} {\bibinfo  {journal} {Comput. Softw. Big Sci.}\ }\textbf {\bibinfo
  {volume} {5}},\ \bibinfo {pages} {2} (\bibinfo {year} {2021})}\BibitemShut
  {NoStop}%
\bibitem [{\citenamefont {Wu}\ \emph {et~al.}(2021)\citenamefont {Wu},
  \citenamefont {Chan}, \citenamefont {Guan}, \citenamefont {Sun},
  \citenamefont {Wang}, \citenamefont {Zhou}, \citenamefont {Livny},
  \citenamefont {Carminati}, \citenamefont {Di~Meglio}, \citenamefont {Li},
  \citenamefont {Lykken}, \citenamefont {Spentzouris}, \citenamefont {Chen},
  \citenamefont {Yoo},\ and\ \citenamefont {Wei}}]{qml4hep2}%
  \BibitemOpen
  \bibfield  {author} {\bibinfo {author} {\bibfnamefont {S.~L.}\ \bibnamefont
  {Wu}}, \bibinfo {author} {\bibfnamefont {J.}~\bibnamefont {Chan}}, \bibinfo
  {author} {\bibfnamefont {W.}~\bibnamefont {Guan}}, \bibinfo {author}
  {\bibfnamefont {S.}~\bibnamefont {Sun}}, \bibinfo {author} {\bibfnamefont
  {A.}~\bibnamefont {Wang}}, \bibinfo {author} {\bibfnamefont {C.}~\bibnamefont
  {Zhou}}, \bibinfo {author} {\bibfnamefont {M.}~\bibnamefont {Livny}},
  \bibinfo {author} {\bibfnamefont {F.}~\bibnamefont {Carminati}}, \bibinfo
  {author} {\bibfnamefont {A.}~\bibnamefont {Di~Meglio}}, \bibinfo {author}
  {\bibfnamefont {A.~C.~Y.}\ \bibnamefont {Li}}, \bibinfo {author}
  {\bibfnamefont {J.~D.}\ \bibnamefont {Lykken}}, \bibinfo {author}
  {\bibfnamefont {P.}~\bibnamefont {Spentzouris}}, \bibinfo {author}
  {\bibfnamefont {S.~Y.-C.}\ \bibnamefont {Chen}}, \bibinfo {author}
  {\bibfnamefont {S.}~\bibnamefont {Yoo}},\ and\ \bibinfo {author}
  {\bibfnamefont {T.-C.}\ \bibnamefont {Wei}},\ }\bibfield  {title} {\bibinfo
  {title} {Application of quantum machine learning using the quantum
  variational classifier method to high energy physics analysis at the lhc on
  ibm quantum computer simulator and hardware with 10 qubits},\ }\href
  {http://iopscience.iop.org/article/10.1088/1361-6471/ac1391} {\bibfield
  {journal} {\bibinfo  {journal} {Journal of Physics G: Nuclear and Particle
  Physics}\ } (\bibinfo {year} {2021})},\ \bibinfo {note} {accepted},\ \Eprint
  {https://arxiv.org/abs/2012.11560} {arXiv:2012.11560 [quant-ph]} \BibitemShut
  {NoStop}%
\bibitem [{\citenamefont {Havlíček}\ \emph {et~al.}(2019)\citenamefont
  {Havlíček}, \citenamefont {Córcoles}, \citenamefont {Temme}, \citenamefont
  {Harrow}, \citenamefont {Kandala}, \citenamefont {Chow},\ and\ \citenamefont
  {Gambetta}}]{qsvmv}%
  \BibitemOpen
  \bibfield  {author} {\bibinfo {author} {\bibfnamefont {V.}~\bibnamefont
  {Havlíček}}, \bibinfo {author} {\bibfnamefont {A.~D.}\ \bibnamefont
  {Córcoles}}, \bibinfo {author} {\bibfnamefont {K.}~\bibnamefont {Temme}},
  \bibinfo {author} {\bibfnamefont {A.~W.}\ \bibnamefont {Harrow}}, \bibinfo
  {author} {\bibfnamefont {A.}~\bibnamefont {Kandala}}, \bibinfo {author}
  {\bibfnamefont {J.~M.}\ \bibnamefont {Chow}},\ and\ \bibinfo {author}
  {\bibfnamefont {J.~M.}\ \bibnamefont {Gambetta}},\ }\bibfield  {title}
  {\bibinfo {title} {{Supervised learning with quantum-enhanced feature
  spaces}},\ }\href@noop {} {\bibfield  {journal} {\bibinfo  {journal}
  {Nature}\ }\textbf {\bibinfo {volume} {567}},\ \bibinfo {pages} {209}
  (\bibinfo {year} {2019})}\BibitemShut {NoStop}%
\bibitem [{\citenamefont {Schuld}\ and\ \citenamefont
  {Killoran}(2019)}]{qsvmm}%
  \BibitemOpen
  \bibfield  {author} {\bibinfo {author} {\bibfnamefont {M.}~\bibnamefont
  {Schuld}}\ and\ \bibinfo {author} {\bibfnamefont {N.}~\bibnamefont
  {Killoran}},\ }\bibfield  {title} {\bibinfo {title} {{Quantum machine
  learning in feature Hilbert spaces}},\ }\href@noop {} {\bibfield  {journal}
  {\bibinfo  {journal} {Physical Review Letters}\ }\textbf {\bibinfo {volume}
  {122}},\ \bibinfo {pages} {040504} (\bibinfo {year} {2019})}\BibitemShut
  {NoStop}%
\bibitem [{\citenamefont {Aaronson}\ and\ \citenamefont
  {Ambainis}(2018)}]{forrelation}%
  \BibitemOpen
  \bibfield  {author} {\bibinfo {author} {\bibfnamefont {S.}~\bibnamefont
  {Aaronson}}\ and\ \bibinfo {author} {\bibfnamefont {A.}~\bibnamefont
  {Ambainis}},\ }\bibfield  {title} {\bibinfo {title} {Forrelation: A problem
  that optimally separates quantum from classical computing},\ }\href@noop {}
  {\bibfield  {journal} {\bibinfo  {journal} {SIAM Journal on Computing}\
  }\textbf {\bibinfo {volume} {47}},\ \bibinfo {pages} {982} (\bibinfo {year}
  {2018})}\BibitemShut {NoStop}%
\bibitem [{\citenamefont {Liu}\ \emph {et~al.}(2020)\citenamefont {Liu},
  \citenamefont {Arunachalam},\ and\ \citenamefont {Temme}}]{qsvm-dlp}%
  \BibitemOpen
  \bibfield  {author} {\bibinfo {author} {\bibfnamefont {Y.}~\bibnamefont
  {Liu}}, \bibinfo {author} {\bibfnamefont {S.}~\bibnamefont {Arunachalam}},\
  and\ \bibinfo {author} {\bibfnamefont {K.}~\bibnamefont {Temme}},\ }\bibfield
   {title} {\bibinfo {title} {A rigorous and robust quantum speed-up in
  supervised machine learning},\ }\href@noop {} {\bibfield  {journal} {\bibinfo
   {journal} {arXiv:2010.02174}\ } (\bibinfo {year} {2020})}\BibitemShut
  {NoStop}%
\bibitem [{\citenamefont {Boser}\ \emph {et~al.}(1992)\citenamefont {Boser},
  \citenamefont {Guyon},\ and\ \citenamefont {Vapnik}}]{svm}%
  \BibitemOpen
  \bibfield  {author} {\bibinfo {author} {\bibfnamefont {B.~E.}\ \bibnamefont
  {Boser}}, \bibinfo {author} {\bibfnamefont {I.~M.}\ \bibnamefont {Guyon}},\
  and\ \bibinfo {author} {\bibfnamefont {V.~N.}\ \bibnamefont {Vapnik}},\
  }\bibfield  {title} {\bibinfo {title} {A training algorithm for optimal
  margin classifiers},\ }in\ \href@noop {} {\emph {\bibinfo {booktitle}
  {Proceedings of the fifth annual workshop on Computational learning
  theory}}}\ (\bibinfo {year} {1992})\ pp.\ \bibinfo {pages}
  {144--152}\BibitemShut {NoStop}%
\bibitem [{\citenamefont {Vapnik}(2013)}]{svm2}%
  \BibitemOpen
  \bibfield  {author} {\bibinfo {author} {\bibfnamefont {V.~N.}\ \bibnamefont
  {Vapnik}},\ }\href@noop {} {\emph {\bibinfo {title} {The Nature of
  Statistical Learning Theory}}}\ (\bibinfo  {publisher} {Springer Science \&
  Business Media},\ \bibinfo {year} {2013})\BibitemShut {NoStop}%
\bibitem [{\citenamefont {Alwall}\ \emph {et~al.}(2014)\citenamefont {Alwall},
  \citenamefont {Frederix}, \citenamefont {Frixione}, \citenamefont {Hirschi},
  \citenamefont {Maltoni}, \citenamefont {Mattelaer}, \citenamefont {Shao},
  \citenamefont {Stelzer}, \citenamefont {Torrielli},\ and\ \citenamefont
  {Zaro}}]{madgraph}%
  \BibitemOpen
  \bibfield  {author} {\bibinfo {author} {\bibfnamefont {J.}~\bibnamefont
  {Alwall}}, \bibinfo {author} {\bibfnamefont {R.}~\bibnamefont {Frederix}},
  \bibinfo {author} {\bibfnamefont {S.}~\bibnamefont {Frixione}}, \bibinfo
  {author} {\bibfnamefont {V.}~\bibnamefont {Hirschi}}, \bibinfo {author}
  {\bibfnamefont {F.}~\bibnamefont {Maltoni}}, \bibinfo {author} {\bibfnamefont
  {O.}~\bibnamefont {Mattelaer}}, \bibinfo {author} {\bibfnamefont {H.-S.}\
  \bibnamefont {Shao}}, \bibinfo {author} {\bibfnamefont {T.}~\bibnamefont
  {Stelzer}}, \bibinfo {author} {\bibfnamefont {P.}~\bibnamefont {Torrielli}},\
  and\ \bibinfo {author} {\bibfnamefont {M.}~\bibnamefont {Zaro}},\ }\bibfield
  {title} {\bibinfo {title} {The automated computation of tree-level and
  next-to-leading order differential cross sections, and their matching to
  parton shower simulations},\ }\href@noop {} {\bibfield  {journal} {\bibinfo
  {journal} {Journal of High Energy Physics}\ }\textbf {\bibinfo {volume}
  {07}},\ \bibinfo {pages} {079} (\bibinfo {year} {2014})}\BibitemShut
  {NoStop}%
\bibitem [{\citenamefont {Sj{\"o}strand}\ \emph {et~al.}(2006)\citenamefont
  {Sj{\"o}strand}, \citenamefont {Mrenna},\ and\ \citenamefont
  {Skands}}]{pythia}%
  \BibitemOpen
  \bibfield  {author} {\bibinfo {author} {\bibfnamefont {T.}~\bibnamefont
  {Sj{\"o}strand}}, \bibinfo {author} {\bibfnamefont {S.}~\bibnamefont
  {Mrenna}},\ and\ \bibinfo {author} {\bibfnamefont {P.}~\bibnamefont
  {Skands}},\ }\bibfield  {title} {\bibinfo {title} {{PYTHIA 6.4 physics and
  manual}},\ }\href@noop {} {\bibfield  {journal} {\bibinfo  {journal} {Journal
  of High Energy Physics}\ }\textbf {\bibinfo {volume} {05}},\ \bibinfo {pages}
  {026} (\bibinfo {year} {2006})}\BibitemShut {NoStop}%
\bibitem [{\citenamefont {De~Favereau}\ \emph {et~al.}(2014)\citenamefont
  {De~Favereau}, \citenamefont {Delaere}, \citenamefont {Demin}, \citenamefont
  {Giammanco}, \citenamefont {Lemaitre}, \citenamefont {Mertens}, \citenamefont
  {Selvaggi}, \citenamefont {Collaboration} \emph {et~al.}}]{delphes}%
  \BibitemOpen
  \bibfield  {author} {\bibinfo {author} {\bibfnamefont {J.}~\bibnamefont
  {De~Favereau}}, \bibinfo {author} {\bibfnamefont {C.}~\bibnamefont
  {Delaere}}, \bibinfo {author} {\bibfnamefont {P.}~\bibnamefont {Demin}},
  \bibinfo {author} {\bibfnamefont {A.}~\bibnamefont {Giammanco}}, \bibinfo
  {author} {\bibfnamefont {V.}~\bibnamefont {Lemaitre}}, \bibinfo {author}
  {\bibfnamefont {A.}~\bibnamefont {Mertens}}, \bibinfo {author} {\bibfnamefont
  {M.}~\bibnamefont {Selvaggi}}, \bibinfo {author} {\bibfnamefont {D.~.}\
  \bibnamefont {Collaboration}}, \emph {et~al.},\ }\bibfield  {title} {\bibinfo
  {title} {{DELPHES 3: a modular framework for fast simulation of a generic
  collider experiment}},\ }\href@noop {} {\bibfield  {journal} {\bibinfo
  {journal} {Journal of High Energy Physics}\ }\textbf {\bibinfo {volume}
  {02}},\ \bibinfo {pages} {057} (\bibinfo {year} {2014})}\BibitemShut
  {NoStop}%
\bibitem [{\citenamefont {Pearson}(1901)}]{pca}%
  \BibitemOpen
  \bibfield  {author} {\bibinfo {author} {\bibfnamefont {K.}~\bibnamefont
  {Pearson}},\ }\bibfield  {title} {\bibinfo {title} {{LIII. On lines and
  planes of closest fit to systems of points in space}},\ }\href@noop {}
  {\bibfield  {journal} {\bibinfo  {journal} {The London, Edinburgh, and Dublin
  Philosophical Magazine and Journal of Science}\ }\textbf {\bibinfo {volume}
  {2}},\ \bibinfo {pages} {559} (\bibinfo {year} {1901})}\BibitemShut {NoStop}%
\bibitem [{\citenamefont {Jolliffe}(2002)}]{pca1}%
  \BibitemOpen
  \bibfield  {author} {\bibinfo {author} {\bibfnamefont {I.}~\bibnamefont
  {Jolliffe}},\ }\href {https://books.google.com/books?id=TtVF-ao4fI8C} {\emph
  {\bibinfo {title} {Principal Component Analysis}}},\ Springer Series in
  Statistics\ (\bibinfo  {publisher} {Springer},\ \bibinfo {year}
  {2002})\BibitemShut {NoStop}%
\bibitem [{\citenamefont {Broughton}\ \emph {et~al.}(2020)\citenamefont
  {Broughton} \emph {et~al.}}]{tfq}%
  \BibitemOpen
  \bibfield  {author} {\bibinfo {author} {\bibfnamefont {M.}~\bibnamefont
  {Broughton}} \emph {et~al.},\ }\href@noop {} {\bibinfo {title} {{TensorFlow
  Quantum: A Software Framework for Quantum Machine Learning}}},\ \bibinfo
  {howpublished} {arXiv:2003.02989} (\bibinfo {year} {2020})\BibitemShut
  {NoStop}%
\bibitem [{\citenamefont {Aleksandrowicz}\ \emph {et~al.}(2019)\citenamefont
  {Aleksandrowicz} \emph {et~al.}}]{qiskit}%
  \BibitemOpen
  \bibfield  {author} {\bibinfo {author} {\bibfnamefont {G.}~\bibnamefont
  {Aleksandrowicz}} \emph {et~al.},\ }\href
  {https://doi.org/10.5281/zenodo.2562110} {\bibinfo {title} {{Qiskit: An
  Open-source Framework for Quantum Computing}}} (\bibinfo {year}
  {2019})\BibitemShut {NoStop}%
\bibitem [{\citenamefont {{Amazon Braket}}(2019)}]{braket}%
  \BibitemOpen
  \bibfield  {author} {\bibinfo {author} {\bibnamefont {{Amazon Braket}}},\
  }\href@noop {} {\bibinfo {title} {https://aws.amazon.com/braket/}} (\bibinfo
  {year} {2019})\BibitemShut {NoStop}%
\bibitem [{\citenamefont {Allen}(1974)}]{cv1}%
  \BibitemOpen
  \bibfield  {author} {\bibinfo {author} {\bibfnamefont {D.~M.}\ \bibnamefont
  {Allen}},\ }\bibfield  {title} {\bibinfo {title} {{The relationship between
  variable selection and data agumentation and a method for prediction}},\
  }\href@noop {} {\bibfield  {journal} {\bibinfo  {journal} {Technometrics}\
  }\textbf {\bibinfo {volume} {16}},\ \bibinfo {pages} {125} (\bibinfo {year}
  {1974})}\BibitemShut {NoStop}%
\bibitem [{\citenamefont {Stone}(1974)}]{cv2}%
  \BibitemOpen
  \bibfield  {author} {\bibinfo {author} {\bibfnamefont {M.}~\bibnamefont
  {Stone}},\ }\bibfield  {title} {\bibinfo {title} {{Cross-validatory choice
  and assessment of statistical predictions}},\ }\href@noop {} {\bibfield
  {journal} {\bibinfo  {journal} {Journal of the Royal Statistical Society,
  Series B (Methodological)}\ }\textbf {\bibinfo {volume} {36}},\ \bibinfo
  {pages} {111} (\bibinfo {year} {1974})}\BibitemShut {NoStop}%
\bibitem [{\citenamefont {Pedregosa}\ \emph {et~al.}(2011)\citenamefont
  {Pedregosa}, \citenamefont {Varoquaux}, \citenamefont {Gramfort},
  \citenamefont {Michel}, \citenamefont {Thirion}, \citenamefont {Grisel},
  \citenamefont {Blondel}, \citenamefont {Prettenhofer}, \citenamefont {Weiss},
  \citenamefont {Dubourg} \emph {et~al.}}]{sklearn}%
  \BibitemOpen
  \bibfield  {author} {\bibinfo {author} {\bibfnamefont {F.}~\bibnamefont
  {Pedregosa}}, \bibinfo {author} {\bibfnamefont {G.}~\bibnamefont
  {Varoquaux}}, \bibinfo {author} {\bibfnamefont {A.}~\bibnamefont {Gramfort}},
  \bibinfo {author} {\bibfnamefont {V.}~\bibnamefont {Michel}}, \bibinfo
  {author} {\bibfnamefont {B.}~\bibnamefont {Thirion}}, \bibinfo {author}
  {\bibfnamefont {O.}~\bibnamefont {Grisel}}, \bibinfo {author} {\bibfnamefont
  {M.}~\bibnamefont {Blondel}}, \bibinfo {author} {\bibfnamefont
  {P.}~\bibnamefont {Prettenhofer}}, \bibinfo {author} {\bibfnamefont
  {R.}~\bibnamefont {Weiss}}, \bibinfo {author} {\bibfnamefont
  {V.}~\bibnamefont {Dubourg}}, \emph {et~al.},\ }\bibfield  {title} {\bibinfo
  {title} {{Scikit-learn: Machine Learning in Python}},\ }\href@noop {}
  {\bibfield  {journal} {\bibinfo  {journal} {Journal of Machine Learning
  Research}\ }\textbf {\bibinfo {volume} {12}},\ \bibinfo {pages} {2825}
  (\bibinfo {year} {2011})}\BibitemShut {NoStop}%
\bibitem [{\citenamefont {Friedman}(2002)}]{bdt1}%
  \BibitemOpen
  \bibfield  {author} {\bibinfo {author} {\bibfnamefont {J.~H.}\ \bibnamefont
  {Friedman}},\ }\bibfield  {title} {\bibinfo {title} {Stochastic gradient
  boosting},\ }\href@noop {} {\bibfield  {journal} {\bibinfo  {journal}
  {Computational Statistics \& Data Analysis}\ }\textbf {\bibinfo {volume}
  {38}},\ \bibinfo {pages} {367} (\bibinfo {year} {2002})}\BibitemShut
  {NoStop}%
\bibitem [{\citenamefont {Hastie}\ \emph {et~al.}(2009)\citenamefont {Hastie},
  \citenamefont {Tibshirani},\ and\ \citenamefont {Friedman}}]{bdt2}%
  \BibitemOpen
  \bibfield  {author} {\bibinfo {author} {\bibfnamefont {T.}~\bibnamefont
  {Hastie}}, \bibinfo {author} {\bibfnamefont {R.}~\bibnamefont {Tibshirani}},\
  and\ \bibinfo {author} {\bibfnamefont {J.}~\bibnamefont {Friedman}},\
  }\href@noop {} {\emph {\bibinfo {title} {The elements of statistical
  learning: data mining, inference, and prediction}}}\ (\bibinfo  {publisher}
  {Springer Science \& Business Media},\ \bibinfo {year} {2009})\BibitemShut
  {NoStop}%
\bibitem [{\citenamefont {Chen}\ and\ \citenamefont
  {Guestrin}(2016)}]{xgboost}%
  \BibitemOpen
  \bibfield  {author} {\bibinfo {author} {\bibfnamefont {T.}~\bibnamefont
  {Chen}}\ and\ \bibinfo {author} {\bibfnamefont {C.}~\bibnamefont
  {Guestrin}},\ }\bibfield  {title} {\bibinfo {title} {{XGBoost: A scalable
  tree boosting system}},\ }in\ \href@noop {} {\emph {\bibinfo {booktitle}
  {Proceedings of the 22nd ACM SIGKDD International Conference on Knowledge
  Discovery and Data Mining}}}\ (\bibinfo {year} {2016})\ pp.\ \bibinfo {pages}
  {785--794}\BibitemShut {NoStop}%
\bibitem [{\citenamefont {{IBM Quantum}}(2021)}]{ibm_web}%
  \BibitemOpen
  \bibfield  {author} {\bibinfo {author} {\bibnamefont {{IBM Quantum}}},\
  }\href@noop {} {\bibinfo {title} {https://quantum-computing.ibm.com/}}
  (\bibinfo {year} {2021})\BibitemShut {NoStop}%
\end{thebibliography}
%

\vspace{0.5cm}
\textbf{Acknowledgements } 
This project is supported in part by the United States Department of Energy, Office of Science, HEP-QIS Research Program, under Award Number DE-SC0020416 and by the Vilas foundation at the University of Wisconsin.
This project is also supported in part by the United States Department of Energy, Office of Science, Office of High Energy Physics program under Award Number DE-SC-0012704 and the Brookhaven National Laboratory LDRD \#20-024.
This research used resources of the Oak Ridge Leadership Computing Facility, which is a United States Department of Energy Office of Science User Facility supported under Contract DE-AC05-00OR22725.
The Wisconsin group would like to thank the ATLAS Collaboration for the inspiration of the LHC flagship analysis used in this publication.

\end{document}